\documentclass[fleqn,usenatbib]{mnras}

\usepackage[T1]{fontenc}
\usepackage{ae,aecompl}
\usepackage{amsfonts,amssymb}
\usepackage{mathrsfs}
\usepackage{color}
\usepackage{enumerate}
\usepackage{booktabs}
\usepackage{soul}

\usepackage{float} 

\usepackage{needspace}
\usepackage{dcolumn}
\usepackage{bm}
\usepackage{graphicx}	
\usepackage{amsmath}	
\usepackage{rotating}
\usepackage{adjustbox}
\usepackage{multirow}
\usepackage{xcolor}
\usepackage{arydshln}

\DeclareMathAlphabet\mathbfcal{OMS}{cmsy}{b}{n}
\definecolor{purple}{RGB}{219, 70, 134}
\definecolor{greenish}{RGB}{90, 219, 154}

\title[Improving sampling and calibration of GRBs as distance indicators]{Improving sampling and calibration of GRBs as distance indicators}

\author[Author et al.]{
Ariadna Montiel,$^{1}$\thanks{E-mail: amontiel@icf.unam.mx}
J. I. Cabrera,$^{2,3}$\thanks{E-mail: jcabrera@ciencias.unam.mx}
Juan Carlos Hidalgo$^{1}$\thanks{E-mail: hidalgo@icf.unam.mx}
\\
$^{1}$Instituto de Ciencias F\'isicas, Universidad Nacional
Aut\'onoma de M\'exico,  62210, Cuernavaca, Morelos, M\'{e}xico\\
$^{2}$Facultad de Ciencias, Universidad Nacional
Aut\'onoma de M\'exico,  04510, Ciudad de M\'exico, M\'{e}xico\\
$^{3}$ Colegio de Ciencias y Humanidades Plantel Sur ,Universidad Nacional
Aut\'onoma de M\'exico,  04500, Ciudad de M\'exico, M\'{e}xico
}

\date{Accepted XXX. Received YYY; in original form ZZZ}

\pubyear{2020}

\begin{document}
\label{firstpage}
\pagerange{\pageref{firstpage}--\pageref{lastpage}}
\maketitle

\begin{abstract}
We present a sample of 74 Gamma-Ray Bursts (GRBs) from the Fermi-GBM catalogue for which we compute the distance moduli and use them to constrain effective dark energy models. To overcome the circularity problem affecting GRBs as distance indicators, we calibrate the Amati relation of our sample with a cosmology-independent technique. Specifically, we use the latest observational Hubble parameter data, including associated systematics, to approximate the cosmic expansion through a Bezier parametric curve. We subsequently obtain the distance moduli of the GRBs and include the data in a suite of recent cosmological observations of the expansion history (Planck Compressed 2018, 2012 BOSS release of BAO data and Pantheon SNIa), to compute Bayesian posterior constraints for the standard cosmological model $\Lambda$CDM, as well as $\omega$CDM, and the CPL parametrization. Throughout the analysis we strive to keep under control the error propagation and limit our GRBs sample to avoid observational bias. As a result, we find no evidence in favour of the alternatives to $\Lambda$CDM model. The latter agrees very well with our calibrated sample of GRBs and presently available luminosity distance probes.

\end{abstract}




\section{\label{sec:level1}Introduction}

In the endeavour to characterise the cosmological expansion, standard candles are a keystone for precise distance determinations. In practice, however, the demand for high precision in many of the distance indicators exposes several sources of bias that prevent astrophysical objects from reaching the status of standard candles. The luminosity of Supernovae of Type Ia  (SNe Ia), for example, is thought to be subject to its environment \citep{Kang:2019azh} and such dependence could only be accounted for through precise observations. See however, the recent analysis of \cite{Rose:2020shp} which reports no evidence for such dependence. 
 
In this sense, any contribution from alternative distance indicators, preferably covering a wide range of redshifts, is key to improve cosmological distance determinations, and ultimately, characterise the Dark Energy component. 

An attractive prospect is the distance modulus of Gamma Ray Bursts (GRBs). For pioneer works see e.g., \citet{Ghirlanda_2004,Liang:2005xb,Firmani:2005gs,Schaefer:2006pa}. Despite the well known dispersion in the luminosity correlations and other sources of uncertainty, GRBs remain good candidates for distance indicators.

GRBs account for the most powerful explosions in the Universe; they are bright enough to be detected up to high redshifts\footnote{The highest redshifts recorded lie at $z=8.2$ (GRB090423) \citep{salvaterra2009,2009Natur.461.1254T} and $z= 9.4$ (GRB090429) \citep{Cucchiara_2011}.}. Therefore, they are often proposed as complementary tools to SNe Ia observations to probe the expansion history of the Universe. 
The prompt emission of GRBs lies mostly in the range from $0.001$ to a few MeV, and lasting from $0.01$ to $1000$ seconds. This property classifies naturally the set of GRBs into two categories, those with $T{90}> 2$ seconds are classified as \textit{long} and are associated to the collapse of certain types of massive stars. On the other hand, the \textit{short} kind is associated to the merger of compact objects \citep{Meszaros:2006rc}. 
Despite several efforts to model the explosion mechanism of the GRBs \citep[e.g.,][]{Mendoza:2007gj,Rivera-Paleo:2016nga,Dainotti:2016qxe,Stratta:2018xza}, there is no single satisfactory explanation and their nature is still not fully understood. In consequence, the distance calibration of GRBs presents more difficulties than that of SNe Ia.  

So far, several methods have been proposed to calibrate GRBs \citep{Liang:2005xb,Firmani:2005gs,Schaefer:2006pa,Liang:2008,doi:10.1063/1.3027949,Wei:2008kq,Kodama08,Wang:2008vja,Liu:2014vda}. Most calibrating methods rely on empirical luminosity correlations found in \textit{long} GRBs. Among the known correlations are those between spectrum lag and isotropic peak luminosity ($\tau_{\rm lag}- L$ relation of  \citet{2000ApJ...534..248N}), the correlation between time variability and isotropic peak luminosity ($V - L$ relation proposed by \citet{Fenimore:2000vs}), a tight correlation between the peak energy of $\nu F_{\nu}$ spectrum and isotropic equivalent energy ($E_{\rm p}-E_{\mathrm{iso}}$ relation of \citet{Amati:2002}), a correlation between the peak energy and the collimation-corrected energy ($E_{\rm p}-E_{\gamma}$ relation proposed by \citet{Ghirlanda:2004me}), the correlation between peak energy and isotropic peak luminosity ($E_{\rm p}-L$ relation of  \citet{2004ApJ...609..935Y}), the correlation between the isotropic peak luminosity, the peak energy and the high signal time-scale, ($L_{\rm iso}-E_{\rm p}-T_{0.45}$ relation by \citet{Firmani2006}) and the correlation between minimum rise time of light curve and isotropic peak luminosity ($\tau_{\rm RT}- L$ relation of \citet{Schaefer:2006pa}). Recently, some other correlations have been reported in the literature \citep{Dainotti:2008vw,2010ApJ...725.2209L,2011MNRAS.410L..47G,2012A&A...538A.134X,L__2012,Yi:2015poa,Liang:2015dua}. For a review and more details about them, see for instance \cite{zhang2018physics}.

In many of the above, the luminosity of GRBs appears correlated with the temporal and spectral properties. While such correlations are not fully understood from first principles, their existence naturally leads to the consideration that GRBs could be used as distance indicators, offering a route to probe the expansion history of the Universe up to $z\gtrsim 9$. Datasets of GRBs distance moduli  are often used (either alone or in combination with other observational data such as SNe Ia), to constrain cosmological parameters  \citep[e.g.][]{Ghirlanda_2004,Firmani:2005gs,Liang:2005xb,Schaefer:2006pa,Liang:2008,doi:10.1063/1.3027949,Wei:2008kq,Wei_2010,Montiel:2011gw,Velten:2012ej,Breton:2013twa,Amati:2013sca,Wang:2015ira}. It is important to note that in the calibration of empirical correlations, two caveats stand out that prevent the improvement of GRBs distance moduli as distance indicators. First is a number of sources of uncertainty in the determination of luminosity parameters. It is known that combining databases from different telescopes may introduce an unknown selection bias due to the different thresholds and spectroscopic sensitivity \citep{Butler2007,Nava:2008hk}.  Additionally, mixing methods for the redshift determination (photometric vs. spectral) also represents a source for uncertainty \citep{Coward:2008te}. 

The purpose of the present work is to present a new and independent dataset of GRBs addressing the above issues and show their usefulness as distance indicators. To overcome the aforementioned shortcomings, we consider data exclusively from a single catalogue (the Fermi-GBM catalogue), which prevents selection biases and other instrument-associated systematics. Moreover, to avoid extra bias, instead of using the complete set of long GRBs with available redshift, we select only those with redshift determined through spectroscopic methods.

An important drawback of the usual GRBs calibration as distance indicators is an inherent circularity problem, since the determination of energy flux typically assumes an underlying cosmological model. Several works have attempted to tackle these circularity problems by adopting model-independent methods to estimate parameters in the calibration (see for instance \cite{Liang:2005sp,Li:2007re,Firmani:2005gs,Li:2007re,Liang:2008,Kodama08,doi:10.1063/1.3027949,Wei:2008kq,Montiel:2011gw,Velten:2012ej,Breton:2013twa}). Here we adopt the strategy  presented by \cite{Amati:2018tso} to overcome this problem. 

This paper presents a sample of 74 Fermi-GRBs, carefully selected to avoid speculation, and consequently large uncertainty, in the determination of luminosity parameters. After listing their spectral properties, we calibrate the set in a model-independent way by employing the Amati relation \citep{Amati:2013sca,Amati:2008hq} which relates the rest frame peak energy of the spectra $E_{\rm p}$ to the isotropic energy emitted $E_{\mathrm{iso}}$. The calibration is performed following the recent work of \cite{Amati:2018tso} where a compilation of 31 measurements of the Hubble parameter was used to fit a Bezier parametric curve in order to obtain the Hubble's rate at arbitrary redshifts without assuming an \textit{a priori} cosmological model. In that paper, the values for the $E_{\mathrm{iso}}$ was determined for 193 GRBs taken from \cite{2017A&A...598A.112D} and references therein. 
In the present work, in contrast, our sample stems from a single catalogue and in the calibration process we take into account the impact of systematic uncertainties of the Hubble parameter measurements (coming from the stellar population synthesis model, and stellar metallicity among others); an aspect overlooked in previous works. Moreover, we take special care in propagating the uncertainties of the best-fit parameters for the Bezier curve as well as for the Amati relation.

The usefulness of our dataset is demonstrated by comparing our results with previous samples through a Bayesian parameter estimation, performed for three popular effective Dark Energy models. Namely, the $\Lambda$ cold dark matter ($\Lambda$CDM), the $\omega$ cold dark matter  ($\omega$CDM), and the Chevalier-Polarski-Linder (CPL) models. The parameter estimation is carried employing the GRB data and the latest compilation of Supernovae Ia data (SNe Ia) \citep{Scolnic:2017caz}, Baryon Acoustic Oscillations (BAO) \citep{Alam:2016hwk,2011MNRAS.416.3017B,Ross:2014qpa}, and Cosmic Microwave Background (CMB) data \citep{Chen:2018dbv}.

The paper is organized as follows. In Sec.~\ref{GRBs} we present in detail our Fermi/GBM GRBs sample; in Sec.~\ref{Calibration} we calibrate the observables of our sample and subsequently the Amati relation. In Sec.~\ref{models} we present the three dark energy (DE) models studied in this work. In Sec.~\ref{methodology} we present the datasets included in our suite of observations to fit parameters of DE models. We discuss our results in Sec.~\ref{Results} and draw conclusions in Sec.~\ref{sec:conclusions}.

\section{Gamma-Ray Bursts observations} 
\label{GRBs}

The GRBs spectrum is mainly, but not exclusively, described in terms of an empirical spectral function, the Band function \citep{Band1993}, which is explicitly  
\begin{equation}
f(E)=
    \begin{cases}
N_{0}\left(\dfrac{E}{100\text{keV}}\right)^{\alpha}\exp\left( -\dfrac{E}{E_{0}}\right)  &    E \leq E_{b}\\
\\
N_{0}\left( \dfrac{E_{b}}{\text{100keV}}\right) ^{(\alpha - \beta)} \exp(\beta - \alpha) \left( \dfrac{E}{100\text{keV}}\right) ^{\beta} &  E > E_{b}  
\end{cases}
\end{equation}
\noindent with $E_{b} = (\alpha-\beta)E_{0}$. This spectrum peaks at (E$_{\rm p,obs}$), which is related to the spectral parameters as  $E_{\rm p,obs}=E_{0}(2+\alpha)$.

Due to the intense radiation emitted in GRBs, it is possible to detect such explosions at high redshift $z$ \citep{salvaterra2009}. The precise determination of $z$ is crucial to infer the distance to the object (luminosity distance), which is necessary to determine the radiated energy ($E_{\rm iso}$) in gamma rays. The redshift can be computed analysing spectral emission or absorption lines of the afterglow spectrum, or by its photometric analysis at lower energy bands (from X-rays to radio), from observations generally performed by auxiliary telescopes \footnote{Another way to determine the redshift is to identify the possible GRB host galaxy and obtain its redshift by standard methods \citep{Hsiao2020}.}.

\subsection{The sample}
\label{subsec:sample}

While the SWIFT satellite has provided the largest number of GRBs with redshift to the existing catalogues, the BAT instrument of this satellite is limited to energies up to $150$ keV \citep{gehrels2004}. This value lies below the average $E_{\rm p,obs}$ of GRBs \citep{kaneko2006}, which prevents the determination of most of the spectral parameters in the Band function or even the cut-off power-law. Consequently, it is impossible to obtain directly the flux and luminosity for many of the GRBs observed by the BAT-SWIFT satellite.

On the other hand, Fermi features two instruments GBM and LAT sensible to energy bands of 8 keV to 40 MeV \citep{meegan2009}, and 100 MeV to 300 GeV \citep{atwood2009}, respectively. 

After SWIFT-BAT, FERMI-GMB is the instrument with the largest count of long GRBs with redshift, with the advantage that GMB allows for the determination of all the spectral parameters in the Band function. This is why we have compiled and reduced a sample of long GRBs with known redshift exclusively from the FERMI-GBM catalogue (recall that the short GRBs do not satisfy \textit{the same} Amati relation as that of the long GRBs and they are also discarded from our sample).

Including long GRBs from other telescopes aside from FERMI-GBM pertains observational biases caused by the different sensitivities in the range of energy intervals and energy fluxes for each instrument \citep{Ghirlanda2008mnras, Nava:2008hk, ghirlanda2012}. Such biases bring uncertainties to the determination of the fluence, and we avoid them by limiting the size of our sample, in contrast with other recent compilations \citep[see e.g.][]{ Amati:2018tso,Demianski:2019vzl}.

For our compilation, FERMI spectral data were taken from the FERMI-GBM catalogue including observations from August 2008 to March 2019 (107 long GRBs) compiled by  \citet{von_Kienlin_2020,gruber2014, vonKienlin2014}. The redshifts were retrieved from the BAT-SWIFT database available at \url{https://swift.gsfc.nasa.gov/archive/grb_table.html/} and the webpage of J. Greiner \url{http://www.mpe.mpg.de/~jcg/grbgen.html}. Noting that some of the GRBs listed in the FERMI-GBM catalogue present no value for the spectral parameters, we reduced the raw data, employing the Gamma Ray Spectral Fitting Package (RMFIT V4.3.2). In particular, we did this for the cases of GRB120712571, GRB180728728, GRB181020792, GRB190114873, GRB190324947.

Since the determination of redshift from photometry is subject to the learning-curve effect, that is, there is a drift in the mean redshift over time as a consequence of different instruments contributing to redshift acquisition (see e.g. \cite{Coward:2008te}), we avoid further bias and discard 8 GRBs with redshift set through such method \citep{nava2912}. Thus, we limit our sample to those GRBs with redshift determined through spectroscopic methods either from the afterglow or from the host galaxy.  
In addition, we discard 7 events (GRB080905705, GRB091020900, GRB101219686, GRB120907017, GRB121211574, GRB130612141 and GRB180205184),  that present a low signal-to-noise or an incomplete light curve which lead to highly uncertain spectral parameters. 

In the same spirit, we also discard 18 GRBs which present significant uncertainties in the spectral parameters, namely $E_{\rm p}$ and $F_{\rm bolo}$, because of their poor contribution to the fitting procedure.

In sum, after selecting objects meeting the above criteria, from the initial sample of 107 GRBs we finally present in Table \ref{parametrosespectrales} a sample of 74 GRBs covering the redshift range $0.117 \leq z \leq 5.283$, together with their spectral parameters and their associated errors. 

In the following section we describe the method to derive the distance moduli for these objects.

\begin{table*}
\scriptsize
\resizebox{\textwidth}{!}{\begin{tabular}{clccccccccccc}
\hline \hline
GRB & z & t$_{90}$ & N$_{0}$ & $\sigma_{\rm N_{0}}$ & $E_{\rm p,obs}$ & $\sigma_{E_{\rm p,obs}}$ & $\alpha$ & $ \sigma_{\alpha_{0}}$ & $\beta$ & $\sigma_{\beta}$ & F$_{\rm bolo}$ & $\sigma_{F_{\rm bolo}}$  \\
 &  & [s] & [ph/cm$^{2}$s keV] & [ph/cm$^{2}$s keV]  & [keV] & [keV] & & & & &[ergs/cm$^{2}$s] & [ergs/cm$^{2}$s] \\  
\hline \hline
*GRB180728728 & 0.1170 & 6.400E+00 & 2.093E-01 & 5.400E-03 & 7.964E+01 & 1.900E+00 & -1.549E+00 & 1.270E-02 & -2.272E+00 & 1.43E-02 & 1.047E-05 &3.378E-07\\
GRB150727793 & 0.3130 & 4.941E+01 & 1.907E-02 & 5.853E+03 & 1.483E+02 & 1.827E+01 & 1.314E-01 & 2.777E-01 & -2.158E+00 & 1.985E-01 &3.366E-07 &1.83E-07 \\
GRB171010792 & 0.3285 & 1.073E+02 & 1.182E-01 & 1.216E-03 & 1.377E+02 & 1.427E+00 & -1.089E+00 & 5.936E-03 & -2.191E+00 & 8.671E-03 &4.746E-06 &7.923E-08 \\
GRB130427324 & 0.3400 & 1.382E+02 & 6.497E-02 & 1.335E-04 & 8.250E+02 & 5.448E+00 & -1.018E+00 & 1.843E-03 & -2.829E+00 & 3.238E-02 &9.345E-06 &7.62E-08 \\
GRB130925173 & 0.3470 & 2.156E+02 & 5.732E-01 & 3.550E-01 & 2.316E+01 & 8.086E-01 & -1.106E-01 & 1.831E-01 & -2.006E+00 & 3.260E-02 &5.969E-07 & 4.123E-07 \\
GRB140606133 & 0.3840 & 2.278E+01 & 9.634E-03 & 5.992E-04 & 5.338E+02 & 1.128E+02 & -1.239E+00 & 4.621E-02 & -2.037E+00 & 4.716E-01 &1.173E-06 &4.062E-07 \\
*GRB190114873 & 0.4250 & 1.164E+02 & 3.796E-02 & 1.600E-04 & 8.997E+02 & 1.570E+01 & -1.072E+00 & 3.690E-03 & -2.586E+00 & 4.180E-02 &5.876E-06 &1.037E-07 \\
GRB091127976 & 0.4903 & 8.701E+00 & 9.052E-02 & 1.361E-02 & 3.546E+01 & 1.550E+00 & -1.254E+00 & 6.618E-02 & -2.216E+00 & 2.009E-02 &1.651E-06 &2.826E-07 \\
GRB090618353 & 0.5400 & 1.124E+02 & 5.788E-02 & 1.287E-03 & 1.490E+02 & 3.286E+00 & -1.114E+00 & 1.308E-02 & -2.239E+00& 2.007E-02 &2.405E-06 &8.453E-08 \\
GRB170607971 & 0.5570 & 2.093E+01 & 2.024E-02 & 2.086E-03 & 1.118E+02 & 9.032E+00 & -1.286E+00 & 5.950E-02 & -2.383E+00 & 0.000E+00 &7.474E-07 & 9.862E-08 \\
GRB141004973 & 0.5730 & 2.560E+00 & 4.976E-01 & 1.829E-01 & 2.781E+01 & 6.640E+00 & 5.630E-02 & 5.375E-01 & -1.891E+00 & 7.405E-02 &6.262E-07 &6.42E-07 \\
GRB130215063 & 0.5970 & 1.437E+02 & 6.217E-03 & 9.168E-04 & 2.099E+02 & 4.231E+01 & -1.059E+00 & 9.077E-02 & -1.615E+00 & 4.154E-02 &7.411E-07 &1.538E-07 \\
GRB131231198 & 0.6420 & 3.123E+01 & 5.624E-02 & 9.278E-04 & 1.781E+02 & 4.031E+00 & -1.218E+00 & 9.639E-03 & -2.305E+00 & 3.373E-02 &2.782E-06 &8.957E-08 \\
GRB161129300 & 0.6450 & 3.610E+01 & 7.848E-03 & 1.238E-03 & 1.464E+02 & 4.261E+01 & -1.037E+00 & 9.784E-02 & -1.954E+00 & 1.622E-01 &4.04E-07 &1.461E-07 \\
GRB180720598 & 0.6540 & 4.890E+01 & 2.926E-02 & 1.914E-04 & 6.360E+02 & 1.543E+01 & -1.171E+00 & 4.805E-03 & -2.490E+00 & 7.095E-02 &3.314E-06 &8.309E-08 \\
GRB080916406 & 0.6890 & 4.634E+01 & 1.501E-02 & 2.834E-03 & 1.057E+02 & 2.045E+01 & -7.807E-01 & 1.063E-01 & -1.774E+00 & 4.619E-02 &6.254E-07 &1.803E-07 \\
GRB111228657 & 0.7163 & 9.984E+01 & 1.089E-02 & 2.170E-03 & 2.651E+01 & 1.252E+00 & -1.582E+00 & 8.062E-02 & -2.443E+00 & 5.903E-02 &3.246E-07 &7.276E-08 \\
GRB140512814 & 0.7250 & 1.480E+02 & 6.663E-03 & 1.395E-04 & 6.910E+02 & 5.824E+01 & -1.225E+00 & 1.754E-02 & -3.540E+00 & 1.616E+01 &7.148E-07 &1.298E-07 \\
GRB160804065 & 0.7360 & 1.316E+02 & 1.165E-02 & 1.888E-03 & 7.139E+01 & 4.175E+00 & -1.03E+00 & 8.786E-02 & -2.819E+00 & 9.034E-01 &1.672E-07 &4.966E-08 \\
GRB090328401 & 0.7360&  6.170E+01&  9.689E-03&  1.951E-04  &  6.505E+02  &  4.377E+01  & -1.083E+00  &  1.657E-02  & -2.390E+00  &  2.323E-01  &  1.183E-06  &  1.010E-07  \\
GRB100816026  &   0.8049  &  2.045E+00  &  6.539E-02  &  7.034E-03  &  1.331E+02  &  7.083E+00  & -3.178E-01  &  7.378E-02  & -2.733E+00  &  2.726E-01  & 8.674E-07  &  1.665E-07  \\
GRB150514774  &   0.8070  &  1.081E+01  &  2.842E-02  &  5.211E-03  &  6.461E+01  &  5.908E+00  & -1.206E+00  &  9.780E-02  & -2.431E+00  &  1.824E-01  &  6.056E-07  &  1.481E-07  \\
GRB151027166  &   0.8100  &  1.234E+02  &  7.868E-03  &  6.011E-04  &  2.014E+02  &  2.447E+01  & -1.247E+00  &  4.736E-02  & -1.955E+00  &  9.463E-02  &  5.688E-07  &  8.607E-08  \\
GRB091003191  &   0.8969  &  2.022E+01  &  2.044E-02  &  6.546E-04  &  3.702E+02  &  2.661E+01  & -1.072E+00  &  2.277E-02  & -2.215E+00  &  1.480E-01  &  1.707E-06  &  1.747E-07  \\
GRB141225959  &   0.9150  &  5.632E+01  &  1.228E-02  &  2.387E-03  &  1.784E+02  &  1.928E+01  & -2.982E-01  &  1.555E-01  & -2.059E+00  &  1.662E-01  &  4.172E-07  &  1.482E-07  \\
GRB140508128  &   1.0270  &  4.429E+01  &  2.331E-02  &  6.635E-04  &  2.574E+02  &  1.212E+01  & -1.182E+00  &  1.890E-02  & -2.319E+00  &  9.382E-02  &  1.440E-06  &  9.273E-08  \\
GRB091208410  &   1.0633  &  1.248E+01  &  5.948E-02  &  2.492E-02  &  4.473E+01  &  1.276E+01  & -6.152E-01  &  2.236E-01  & -1.923E+00  &  4.911E-02  &  5.556E-07  &  3.398E-07  \\
GRB180620660  &   1.1175  &  4.672E+01  &  1.042E-02  &  1.819E-03  &  1.756E+02  &  4.979E+01  & -1.206E+00  &  1.156E-01  & -1.660E+00  &  3.474E-02  &  1.018E-06  &  2.336E-07  \\
GRB160509374  &   1.1700  &  3.697E+02  &  2.201E-02  &  3.058E-04  &  3.552E+02  &  9.877E+00  & -1.015E+00  &  9.913E-03  & -2.232E+00  &  4.756E-02  &  1.740E-06  &  6.529E-08  \\
*GRB190324947  &   1.1715  &  2.688E+01  &  2.640E-02  &  2.180E-03  &  1.297E+02  &  8.230E+00  & -9.824E-01  &  4.490E-02  & -2.365E+00  &  1.350E-01  &  7.756E-07  &  1.055E-07  \\
 GRB140213807  &   1.2076  &  1.862E+01  &  3.202E-02  &  2.246E-03  &  8.615E+01  &  4.100E+00  & -1.126E+00  &  3.499E-02  & -2.252E+00  &  5.515E-02  &  8.495E-07  &  8.243E-08  \\
GRB090926914  &   1.2400  &  6.400E+01  &  8.279E-02  &  1.367E-02  &  8.234E+01  &  2.561E+00  &  2.346E-01  &  1.057E-01  & -3.343E+00  &  4.482E-01  &  2.152E-07  &  4.965E-08  \\
GRB130420313  &   1.2970  &  1.050E+02  &  2.877E-02  &  8.411E-03  &  5.275E+01  &  3.718E+00  & -9.366E-01  &  1.738E-01  & -2.921E+00  &  4.152E-01  &  2.583E-07  &  9.624E-08  \\
 GRB140801792  &   1.3200  &  7.168E+00  &  1.046E-01  &  6.316E-03  &  1.194E+02  &  2.638E+00  & -3.846E-01  &  3.932E-02  & -3.853E+00  &  1.200E+00  &  1.134E-06  &  2.504E-07  \\
GRB100414097  &   1.3680  &  2.650E+01  &  2.440E-02  &  2.832E-04  &  6.635E+02  &  1.537E+01  & -6.242E-01  &  1.384E-02  & -3.534E+00  &  1.245E+00  &  3.581E-06  &  7.504E-07  \\
GRB100615083  &   1.3980  &  3.738E+01  &  2.607E-02  &  8.130E-03  &  5.355E+01  &  7.505E+00  & -9.064E-01  &  1.560E-01  & -1.803E+00  &  3.070E-02  &  6.146E-07  &  2.330E-07  \\
GRB160625945  &   1.4060  &  4.534E+02  &  2.479E-02  &  1.531E-04  &  4.715E+02  &  6.445E+00  & -9.341E-01  &  4.352E-03  & -2.182E+00  &  2.018E-02  &  2.558E-06  &  4.455E-08  \\
GRB100814160  &   1.4400  &  1.505E+02  &  2.373E-02  &  3.611E-03  &  1.277E+02  &  8.730E+00  & -2.419E-01  &  1.001E-01  & -2.437E+00  &  2.555E-01  &  3.307E-07  &  8.634E-08  \\
GRB180314030  &   1.4450  &  2.202E+01  &  7.343E-02  &  9.386E-03  &  1.030E+02  &  4.485E+00  & -4.038E-01  &  7.725E-02  & -3.356E+00  &  1.473E+00  &  6.711E-07  &  1.678E-07  \\
GRB110213220  &   1.4600  &  3.430E+01  &  8.441E-03  &  7.374E-04  &  1.126E+02  &  1.204E+01  & -1.563E+00  &  4.812E-02  & -4.870E+00  &  0.000E+00  &  4.099E-07  &  5.837E-08  \\
GRB161117066  &   1.5490  &  1.222E+02  &  2.858E-02  &  2.805E-03  &  8.067E+01  &  3.045E+00  & -8.111E-01  &  5.158E-02  & -3.023E+00  &  5.141E-01  &  3.354E-07  &  4.416E-08  \\
GRB100728095  &   1.5670  &  1.654E+02  &  1.973E-02  &  5.593E-04  &  2.539E+02  &  6.569E+00  & -5.097E-01  &  2.151E-02  & -2.542E+00  &  9.954E-02  &  8.905E-07  &  5.625E-08  \\
GRB100906576  &   1.7270  &  1.106E+02  &  2.684E-02  &  1.039E-02  &  7.491E+01  &  2.427E+01  & -9.263E-01  &  2.344E-01  & -1.861E+00  &  1.001E-01  &  7.480E-07  &  4.083E-07  \\
GRB120119170  &   1.7280  &  5.530E+01  &  2.270E-02  &  1.220E-03  &  1.828E+02  &  1.045E+01  & -9.550E-01  &  3.240E-02  & -2.366E+00  &  1.603E-01  &  9.094E-07  &  9.610E-08  \\
GRB150314205  &   1.7580  &  1.069E+01  &  9.181E-02  &  1.649E-03  &  3.472E+02  &  7.897E+00  & -6.792E-01  &  1.382E-02  & -2.601E+00  &  1.021E-01  &  6.361E-06  &  2.866E-07  \\
GRB120326056  &   1.7980  &  1.178E+01  &  6.022E-02  &  2.483E-02  &  4.431E+01  &  5.587E+00  & -6.790E-01  &  2.284E-01  & -2.335E+00  &  1.347E-01  &  3.562E-07  &  1.898E-07  \\
 GRB131011741  &   1.8740  &  7.706E+01  &  8.097E-03  &  8.791E-04  &  2.176E+02  &  4.088E+01  & -8.778E-01  &  7.417E-02  & -2.085E+00  &  2.288E-01  &  4.283E-07  &  1.178E-07  \\
GRB170705115  &   2.0100  &  2.278E+01  &  2.682E-02  &  3.314E-03  &  9.788E+01  &  7.644E+00  & -9.911E-01  &  6.923E-02  & -2.303E+00  &  1.083E-01  &  6.422E-07  &  1.119E-07  \\ 
GRB161017745  &   2.0127  &  3.789E+01  &  9.605E-03  &  1.253E-03  &  2.386E+02  &  4.077E+01  & -1.030E+00  &  1.010E-01  & -2.371E+00  &  7.765E-01  &  5.130E-07  &  1.519E-07  \\
GRB140620219  &   2.0400  &  4.583E+01  &  2.570E-02  &  1.044E-02  &  6.948E+01  &  1.072E+01  & -8.497E-01  &  1.905E-01  & -2.092E+00  &  8.071E-02  &  4.304E-07  &  2.091E-07  \\
GRB150403913  &   2.0600  &  2.227E+01  &  2.561E-02  &  6.409E-04  &  4.287E+02  &  2.106E+01  & -8.733E-01  &  1.830E-02  & -2.108E+00  &  5.752E-02  &  2.484E-06  &  1.619E-07  \\
GRB090926181  &   2.1062  &  1.376E+01  &  6.448E-02  &  7.427E-04  &  3.338E+02  &  5.839E+00  & -8.480E-01  &  8.619E-03  & -2.378E+00  &  4.558E-02  &  4.542E-06  &  1.285E-07  \\
GRB120624933  &   2.1974  &  2.714E+02  &  9.779E-03  &  1.307E-04  &  6.376E+02  &  2.451E+01  & -9.163E-01  &  1.227E-02  & -2.217E+00  &  6.592E-02  &  1.285E-06  &  6.136E-08  \\
GRB121128212  &   2.2000  &  1.734E+01  &  6.344E-02  &  1.620E-02  &  6.008E+01  &  3.849E+00  & -6.837E-01  &  1.195E-01  & -2.424E+00  &  9.205E-02  &  5.345E-07  &  1.644E-07  \\
GRB081221681  &   2.2600  &  2.970E+01  &  6.972E-02  &  2.871E-03  &  8.691E+01  &  1.328E+00  & -8.387E-01  &  2.245E-02  & -3.675E+00  &  4.704E-01  &  1.053E-06  &  2.001E-07  \\
GRB141028455  &   2.3300  &  3.149E+01  &  1.804E-02  &  7.357E-04  &  2.931E+02  &  1.798E+01  & -8.420E-01  &  2.806E-02  & -1.966E+00  &  5.182E-02  &  1.319E-06  &  1.135E-07  \\
GRB130518580  &   2.4900  &  4.858E+01  &  2.066E-02  &  4.170E-04  &  3.811E+02  &  1.457E+01  & -8.629E-01  &  1.561E-02  & -2.181E+00  &  6.691E-02  &  1.746E-06  &  9.416E-08  \\
 GRB081121858  &   2.5120  &  4.198E+01  &  3.519E-02  &  5.452E-03  &  1.609E+02  &  1.445E+01  & -4.351E-01  &  1.148E-01  & -2.096E+00  &  9.481E-02  &  1.029E-06  &  2.568E-07  \\
GRB170214649  &   2.5300  &  1.229E+02  &  1.981E-02  &  2.047E-04  &  4.814E+02  &  1.123E+01  & -9.788E-01  &  8.669E-03  & -2.512E+00  &  1.021E-01  &  1.944E-06  &  7.330E-08  \\
GRB120811649  &   2.6710  &  1.434E+01  &  4.445E-02  &  1.558E-02  &  5.539E+01  &  3.931E+00  & -7.028E-01  &  2.255E-01  & -2.839E+00  &  3.367E-01  &  3.252E-07  &  1.581E-07  \\
GRB140206304  &   2.7400  &  2.726E+01  &  1.006E-01  &  1.408E-02  &  1.212E+02  &  5.826E+00  &  5.490E-02  &  9.329E-02  & -2.416E+00  &  9.637E-02  &  1.091E-06  &  2.308E-07  \\
 GRB081222204  &   2.7700  &  1.888E+01  &  2.546E-02  &  1.874E-03  &  1.472E+02  &  8.431E+00  & -8.444E-01  &  4.478E-02  & -2.300E+00  &  1.188E-01  &  7.994E-07  &  9.551E-08  \\
 GRB110731465  &   2.8300  &  7.485E+00  &  3.837E-02  &  1.457E-03  &  3.222E+02  &  1.692E+01  & -8.686E-01  &  3.117E-02  & -2.436E+00  &  2.741E-01  &  2.652E-06  &  2.598E-07  \\
*GRB181020792  &   2.9380  &  1.510E+01  &  4.340E-02  &  2.350E-03  &  2.694E+02  &  1.350E+01  & -3.552E-01  &  3.960E-02  & -1.778E+00  &  2.270E-02  &  3.062E-06  &  3.062E-07  \\
 GRB140703026  &   3.1400  &  8.397E+01  &  4.179E-03  &  4.153E-04  &  2.089E+02  &  3.473E+01  & -1.267E+00  &  5.914E-02  & -2.681E+00  &  9.169E-01  &  2.319E-07  &  7.141E-08  \\
GRB140423356  &   3.2600  &  9.523E+01  &  1.305E-02  &  2.538E-03  &  1.162E+02  &  1.589E+01  & -5.542E-01  &  1.155E-01  & -1.786E+00  &  4.965E-02  &  3.618E-07  &  1.002E-07  \\
GRB140808038  &   3.2900  &  4.477E+00  &  4.924E-02  &  7.458E-03  &  1.174E+02  &  6.459E+00  & -4.223E-01  &  1.004E-01  & -2.868E+00  &  4.827E-01  &  7.875E-07  &  2.411E-07  \\
GRB110818860  &   3.3600  &  6.707E+01  &  4.822E-03  &  1.088E-03  &  1.822E+02  &  5.789E+01  & -1.112E+00  &  1.460E-01  & -1.765E+00  &  1.571E-01  &  3.033E-07  &  1.053E-07  \\
 GRB170405777  &   3.5100  &  7.859E+01  &  2.005E-02  &  5.696E-04  &  2.670E+02  &  9.288E+00  & -7.993E-01  &  1.996E-02  & -2.354E+00  &  8.888E-02  &  1.170E-06  &  7.549E-08  \\  
GRB090323002  &   3.5700  &  1.339E+02  &  1.101E-02  &  1.838E-04  &  4.536E+02  &  2.360E+01  & -1.183E+00  &  1.145E-02  & -2.354E+00  &  1.470E-01  &  1.013E-06  &  6.838E-08  \\
GRB120909070  &   3.9300  &  1.121E+02  &  6.395E-03  &  4.727E-04  &  1.996E+02  &  2.428E+01  & -8.436E-01  &  5.132E-02  & -1.934E+00  &  7.372E-02  &  3.183E-07  &  4.987E-08  \\
GRB090516353  &   4.1090  &  1.231E+02  &  4.287E-03  &  4.314E-04  &  1.421E+02  &  2.645E+01  & -1.517E+00  &  5.265E-02  & -2.304E+00  &  2.701E-01  &  2.581E-07  &  4.088E-08  \\
GRB120712571  &   4.1745  &  2.253E+01  &  2.395E-02  &  1.140E-02  &  1.194E+02  &  1.610E+01  &  1.437E-01  &  3.020E-01  & -2.163E+00  &  2.100E-01  &  2.732E-07  &  1.790E-07  \\
GRB140304557  &   5.2830  &  3.123E+01  &  9.363E-03  &  2.406E-03  &  1.224E+02  &  3.144E+01  & -7.893E-01  &  1.765E-01  & -2.429E+00  &  6.779E-01  &  2.493E-07  &  1.414E-07  \\
\hline
\end{tabular}}
\caption{Spectral parameters for the employed GRBs taken from the GBM-FERMI catalogue. The (*) represents the GRBs that we processed. Columns are: name, redshift, t$_{90}$, spectral normalization, the standard deviation for the spectral normalization, observed peak energy, standard deviation for the observed peak energy, spectral index of low energy, standard deviation for the spectral index of low energy, spectral index of high energy, standard deviation for the spectral index of high energy, bolometric fluence and the standard deviation for the bolometric fluence.
}\label{parametrosespectrales}
\end{table*}


\section{Calibration}
\label{Calibration}

\begin{figure}
   \centering
   \includegraphics[width=3.2 in]{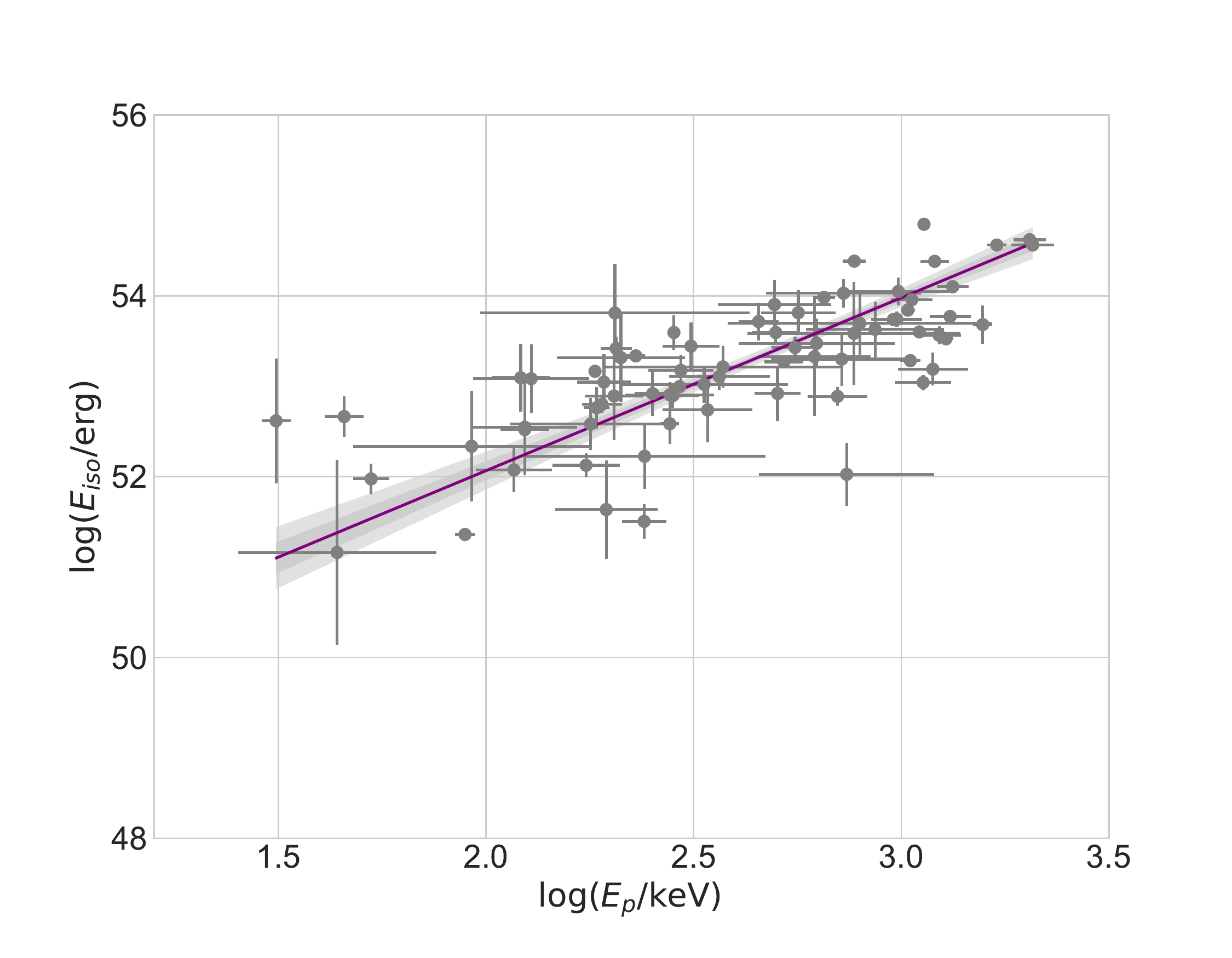}
    \caption{Cosmology-dependent fit of the Amati correlation obtained with the data in Table \ref{parametrosespectrales} assuming a standard flat cosmology with parameter values $h=0.674 \pm 0.005$, and $\Omega_{m}=0.315 \pm 0.007$ \citep{Aghanim:2018eyx}. Note we have performed the corresponding error propagation to calculate the uncertainties for $E_{\rm iso}$. In this case,  the parameters of the Amati correlation in Eq.~\eqref{eq:Amati81} take values $A =1.9099 \pm 0.2412$  and $B = 48.2477 \pm 0.6591$.  The Spearman's rank coefficient of the correlation is $\rho= 0.7592$.}
   \label{ref:EisoEp_LCDM}
\end{figure}

We followed the model-independent calibration recently proposed by \cite{Amati:2018tso}. We thus apply the empirical relation $E_{\rm p}-E_{\mathrm{iso}}$ of \cite{Amati:2002,Amati:2008hq}, that connects $E_{\rm p} = E_{\mathrm{p,obs}} (1+z)$ with the isotropic equivalent energy,  
\begin{equation}
    E_{\mathrm{iso}}(z)=4\pi d^2_{L}(z)S_{\mathrm{bolo}}(1+z)^{-1},
    \label{Ec:Eiso}
\end{equation}
where $S_{\mathrm{bolo}}$ is the bolometric fluence of gamma rays in the GRB at redshift $z$. The factor $(1+z)^{-1}$ transforms the observed GRB duration to the source cosmological rest-frame, and $d_L(z)$ is the luminosity distance of the GRB given by 
\begin{equation}
    d_{L}(\Omega_K,z)= \frac{c}{H_0} \frac{(1+z)}{\sqrt{|\Omega_K|}} \mathrm{sinn} \left[ \sqrt{|\Omega_K |} \int_{0}^{z} \frac{H_0 dz'}{H(z')} \right].
    \label{Ec:d_L}
\end{equation}
In this last expression $\Omega_K$ is the present curvature density defined as $\Omega_K\equiv -K/H_0^2 a^2 $. Also, the symbol $\mathrm{sinn}(x)$ stands for $\sinh{(x)}$ (if $\Omega_K>0$), $\sin{(x)}$ (if $\Omega_K<0$) or just $x$ (if $\Omega_K=0$). 
From this equation it is evident that the calibration of GRBs depends on the cosmological model through the expansion history $H(z)$. In fact a good fit can be obtained when a cosmological model is assumed \textit{a priori} (see Fig.~\ref{ref:EisoEp_LCDM}) although this is the cause of the circularity problem we want to avoid. 

Since the most recent results from the Planck satellite infer a parameter value $\Omega_K= 0.001 \pm 0.002$ \citep{Aghanim:2018eyx}, the analyses in this paper assume $\Omega_K=0$.

Following \cite{Amati:2018tso}, we construct a B\'ezier parametric curve of degree $n$ given by
\begin{align}
        &H_{n}(z)=\sum_{d=0}^{n} \beta_{d}h_{n}^{d}(z), &h_{n}^{d}\equiv \frac{n!(z/z_{m})^{d}}{d!(n-d)!}\left(1-\frac{z}{z_{m}} \right)^{n-d},
    	\label{Ec:Bezier}
    \end{align}
where $\beta_d$ are coefficients of the linear combination of Bernstein basis polynomials $h_{n}^{d}(z)$, positive in the range $0 \leq z/z_m \leq 1$, with $z_m$ the maximum $z$ in the dataset employed. Specifically, we use Hubble parameter data reported by \cite{Capozziello:2017nbu}, which comes from the Cosmic Chronometers (CC) approach \citep{Jimenez_2002,10.1093/mnrasl/slv037}, to build a B\'ezier curve of degree $n=2$ in order to obtain a monotonic growing function in such way that the Hubble constant, $H_0$, can be identified with the parameter $\beta_0$ by setting $d=0$ and $z=0$.

 Cosmic Chronometers is a cosmology-independent technique to measure the Hubble parameter as a function of redshift, $H(z)$, from the differential evolution of massive and passive early-type galaxies. The strength of the CC approach is to provide a direct estimation of the expansion history of the Universe without relying on any cosmological assumption; however, the systematic uncertainties must be carefully taken into account. Such uncertainties are mostly attributed to four sources: the dependence on the stellar population synthesis (SPS) model used to calibrate the measurement, the dependence on the estimate of the stellar metallicity of the population, the dependence on the assumed model of star formation history (SFH), and finally the influence of a possible residual star formation due to a young sub-dominant component underlying the selected sample \citep{Moresco:2020fbm}.  
 An initial estimation of the impact of systematic uncertainties was presented by \citet{Moresco2012b} and \citet{Moresco:2016mzx}, where the impact of the SFH assumption in the method was estimated to be $2.5 \%$ on average. A more recent work showed that it is possible to minimize the impact of a recent burst of star formation through a careful selection of purely passively evolving galaxies from the optical spectra alone \citep{Moresco_2018}. 
 
 Here we follow the latest analysis of \cite{Moresco:2020fbm}, where the systematic uncertainties due to the choice of the stellar population synthesis model (i.e., stellar physics models, along with an adopted stellar library, initial mass function, etc.), and metallicity were estimated. The authors showed that the most relevant effects come from the SPS model, the stellar libraries and metallicity. In order to assess the impact of these systematic uncertainties in our calibration (and in our analysis of the cosmological models), we added to the measurement error of the Hubble parameter dataset we take from \cite{Capozziello:2017nbu}, in quadrature, the maximum bias reported in Table 4 of \cite{Moresco:2020fbm} due to the stellar library ($7.40 \%$), SPS model ($15.86 \%$) and due to $5\%$ error on metallicity ($6.97 \%$), that is, 
 \begin{equation}
 \sigma_{\rm syst }= \pm 7.40 \% ({\rm StellarLib}) \pm 15.86 \% ({\rm SPS}) \pm 6.97 \% ({\rm met}).
 \end{equation}
 
For consistency, and given that the analysis of \cite{Moresco:2020fbm} was carried out through simulations in the redshift range $0<z<1.5$, we limit our CC Hubble data to this range and instead of using the 31 measurements of $H(z)$ reported by \cite{Capozziello:2017nbu}, we work with a sub-sample of 28 measurements of Hubble parameter with maximum redshift $z=1.43$.

\begin{figure*}
\begin{center}
\includegraphics[width=3.in]{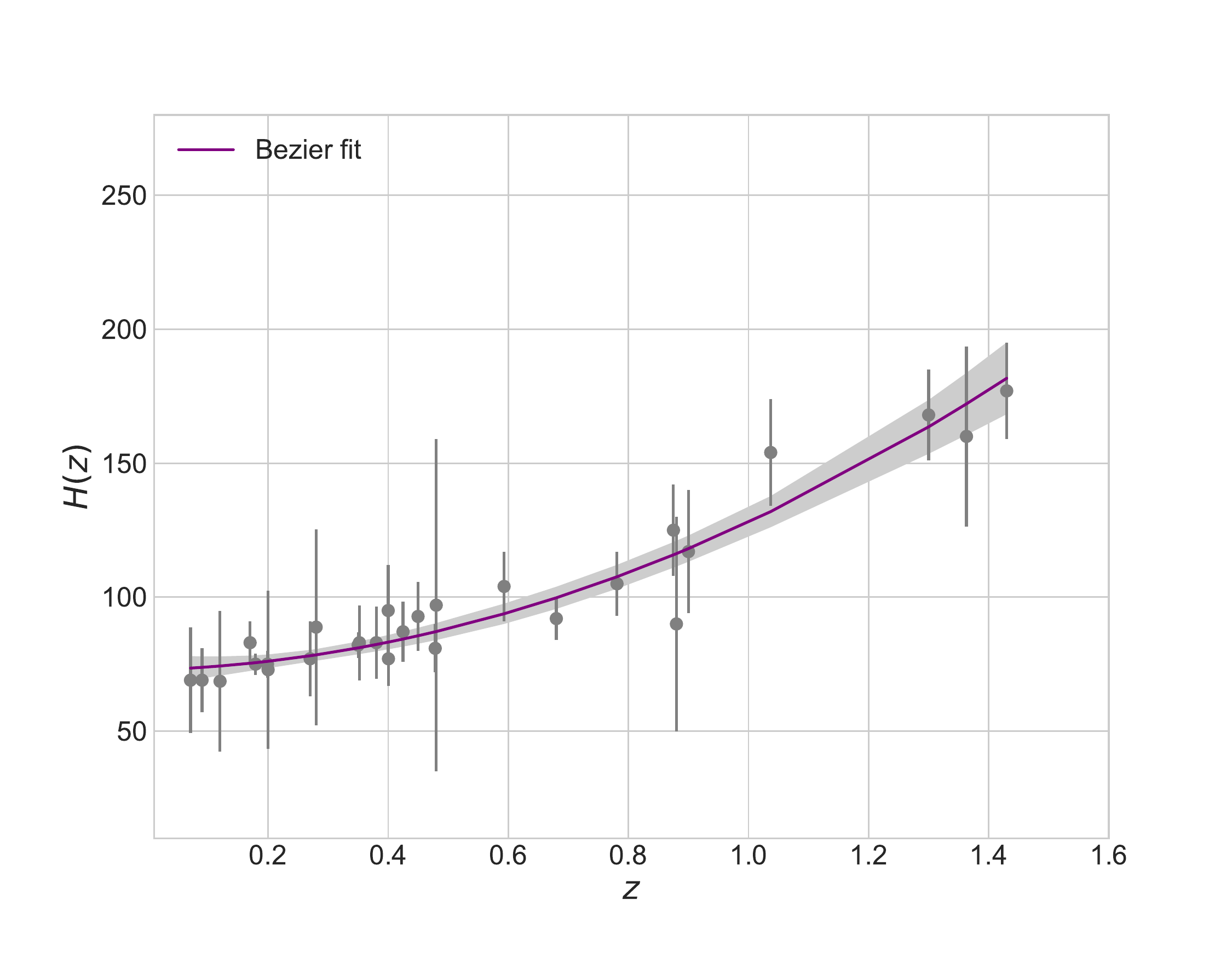}
\includegraphics[width=3.in]{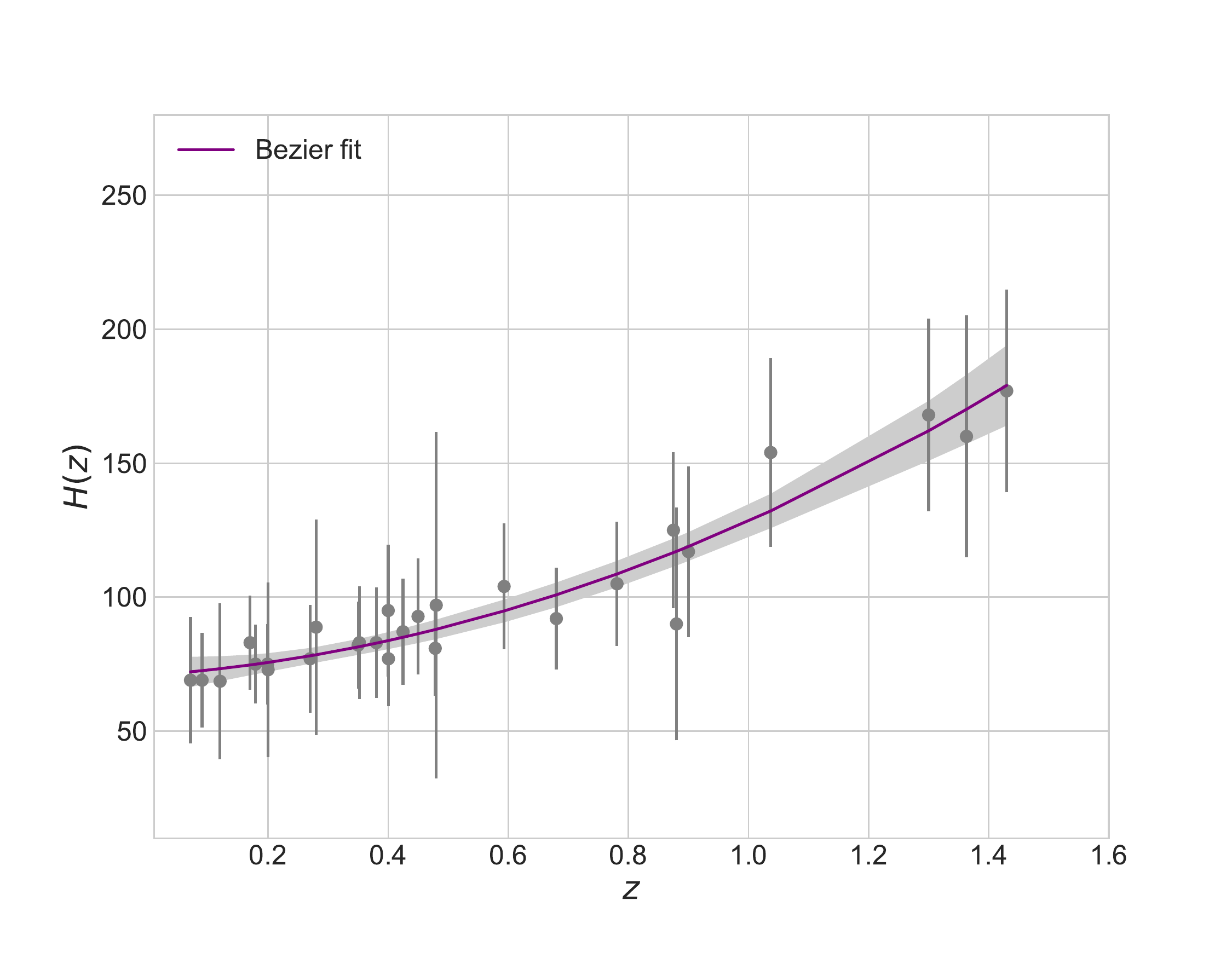}
\end{center}
\caption{The 28 Hubble parameter datapoints with the fit to the B\'ezier curve and its respective 1$\sigma$ confidence region (shaded region). The left panel shows data with the usual errors associated to the differential age \citep{Jimenez_2002} but without extra systematic errors. The right panel includes the maximum uncertainty at 1-$\sigma$ from the recent treatment of uncertainties from SPS modelling, stellar library, and stellar metallicity \citep{Moresco:2020fbm}.
}
\label{fig:BezierCurve}
\end{figure*}

By employing the sub-sample of 28 measurements of the Hubble parameter, we performed a non-linear least-squares minimization by using the Python software package \textsc{lmfit} \citep{newville_matthew_2014_11813} with two error sets. In the first case, presented only for reference, we use the data without adding the bias due to the systematic uncertainties. The second case takes into account the contribution of the maximum systematic error described above. We adopt this second case to perform the subsequent steps of our analysis. The best-fit parameters obtained for the B\'ezier fit with $n=2$ in both cases are
\begin{equation}
    H_{2}(z)=\beta_{0} h_{2}^{0}(z)+\beta_{1} h_{2}^{1}(z)+\beta_{2} h_{2}^{2}(z),
    \label{Ec:Bezier-n2}
\end{equation}  
where for each case,
\begin{eqnarray}
\beta_0^{(I)} =H_0^{(I)}=72.81, & \beta_1^{(I)}=77.46,   &\beta_2^{(I)}=181.71,  \notag\\
\beta_0^{(II)} = H_0^{(II)}=70.81,  &\beta_1^{(II)}=81.99, &\beta_2^{(II)}=179.02.
\label{betas:2cases}
\end{eqnarray}
The corresponding covariance matrices are, 
\[
\mathbf{cov}^{(I)}=
  \begin{bmatrix}
8.58 & -15.20 & 6.91\\
-15.20 & 39.39 & -23.17\\
6.91 & -23.17 &  43.14\\
  \end{bmatrix},
\]

\[
\mathbf{cov}^{(II)}=
  \begin{bmatrix}
12.99 & -20.07 & 8.88\\
-20.07 & 50.18 &  -29.92\\
8.88 & -29.926 &  53.76\\
  \end{bmatrix},
\]

\noindent with the associated errors for the parameters in \eqref{betas:2cases} encoded in the diagonal terms of each matrix.
The best-fit with its 1$\sigma$ confidence region for both cases are shown in each panel of Fig. \ref{fig:BezierCurve}.




\begin{figure}
\begin{center}
\includegraphics[width=3.in]{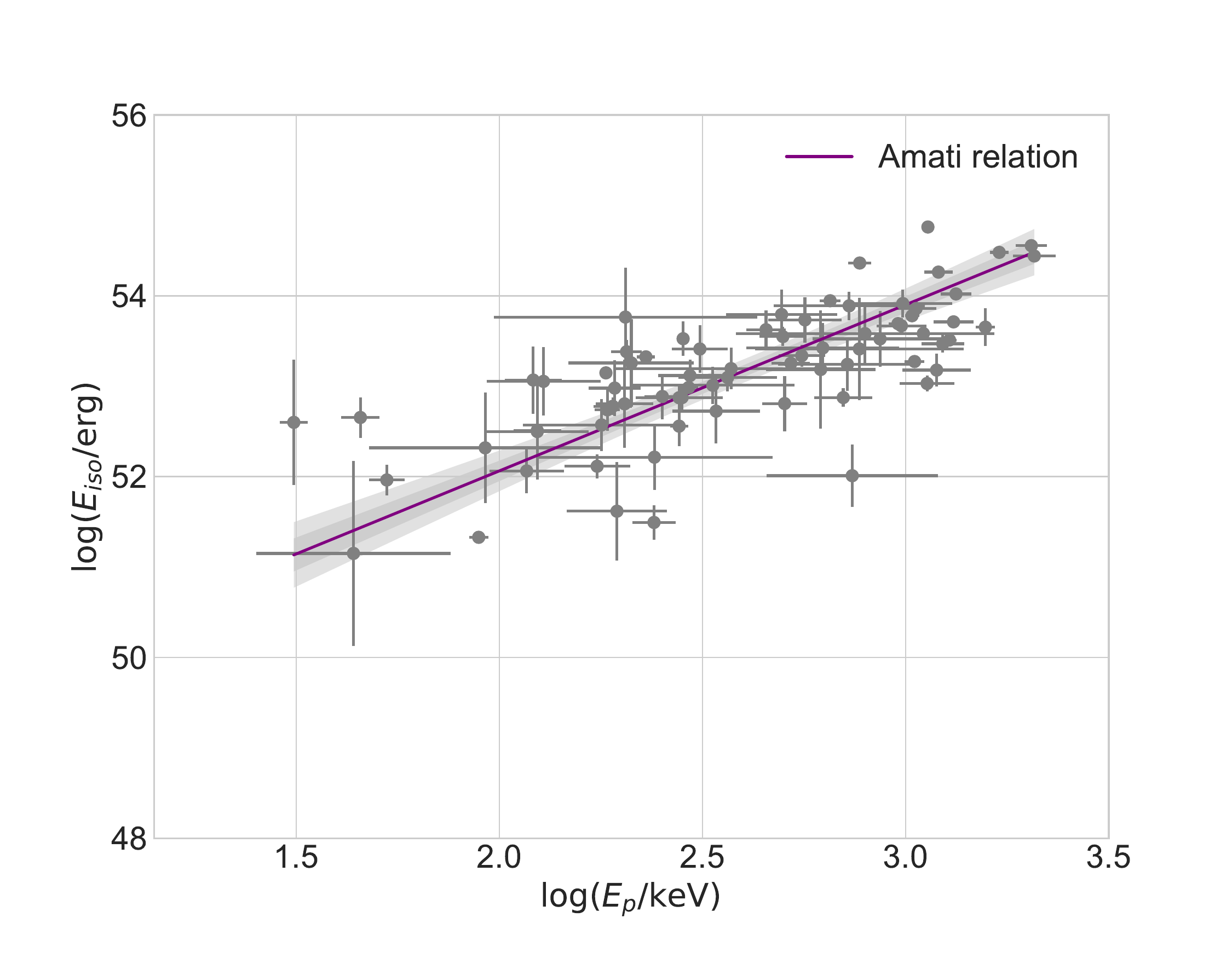}
\end{center}
\caption{74 GRBs between the redshift range $0.117 \leq z \leq 5.283$. The best-fit calibration, Amati relation, is the straight line in purple. In this case the $E_{\mathrm{iso}}$ comes from the CC Hubble data that take into account the maximum extra bias. The Spearman's rank coefficient of the correlation is $\rho= 0.7623$.}
\label{fig:Amati_relation}
\end{figure}

The next step consists of extrapolating the function $H_2(z)$ to redshift $z>z_m$ and construct the luminosity distance $d^{\mathrm{cal}}_{L}(z)$ for a flat cosmology, that is
\begin{equation}
    d^{\mathrm{cal}}_{L}(z)=c(1+z)\int_{0}^{z} \frac{dz'}{H_2(z')},
    \label{Ec:dLcal}
\end{equation}
and subsequently the isotropic energy ${E^{\mathrm{cal}}_{\mathrm{iso}}=4\pi (d^{\mathrm{cal}}_{L}(z))^2S_{\mathrm{bolo}}(1+z)^{-1}} $. In order to obtain the corresponding errors $\sigma E^{\mathrm{cal}}_{\mathrm{iso}}$, the associated error to $\sigma^{\mathrm{cal}}_{d_L}$ is calculated by taking into account the correlations between the parameters $\beta$'s, see Eq. \ref{Ec:Bezier-n2}, besides of the GRBs systematics on the observables, $S_{\mathrm{bolo}}$. In Fig. \ref{fig:Amati_relation}, we show the corresponding $E_{\rm p}- E^{\mathrm{cal}}_{\mathrm{iso}}$ distribution, with $E^{\mathrm{cal}}_{\mathrm{iso}}$ coming from the $H(z)$ data that takes into account the maximum bias due to systematic uncertainties.  


To calibrate the Amati relation for our sample of 74 Fermi-GRBs, we employed the bivariate correlated errors and intrinsic scatter (BCES) method of the Python module by \cite{2012Sci...338.1445N}. This fitting method performs a robust linear regression, following \cite{Akritas:1996kz}, which is useful when it is unclear which variable should be treated as the independent one and which as the dependent one. Moreover, the method takes into account the possible intrinsic scatter of the data with respect to the linear fit and the measurement errors in both variables.

Following this method, the best fit for the Amati relation 
\begin{equation}
    \log\left(\frac{E^{\mathrm{cal}}_{\mathrm{iso}}}{\mathrm{erg}}\right)= A \log \left(\frac{E_{\rm p}}{\mathrm{keV}}\right) +B,
    \label{eq:Amati81}
\end{equation}
is obtained for $A=1.8355\pm 0.2403$, $B=48.3934 \pm0.6551$, and the covariance matrix given by\footnote{ This is in good agreement with the best fit for the Amati relation using the CC Hubble dataset without the extra bias due to systematic uncertainties (case I, where $A=1.8356\pm 0.2395$, $B=48.3885 \pm 0.6521$) which indicates that up to  this point, we find a negligible impact of systematic uncertainties added to the Hubble parameter measurements.}
\begin{equation}
\textbf{cov} = \left[
\begin{array}{cccc}
0.0577 & -0.1568 \\
-0.1568 & 0.4292 \\
\end{array}
\right].
\end{equation}

It should be noted that while we used the BCES-orthogonal regression algorithm to perform our data-fitting, there are other methods of fitting Eq. \eqref{eq:Amati81} to our sample of GRBs. For example, we also considered the \textsc{scipy.odr} package, translated from the \textsc{Fortran-77 odrpack} by \cite{10.1145/76909.76913}. \textsc{odrpack} is a weighted orthogonal distance regression function which takes into account errors in both variables by minimizing the weighted orthogonal distance between the observations and model. With this algorithm, the best fit to the Amati relation was estimated to be $A=1.9287 \pm 0.1862$, $B=48.0725\pm 0.5321$. While this estimate agrees with our results from BCES-orthogonal method and with the results of \cite{Amati:2018tso}, some issues with  the values of uncertainties on the parameters have been noted in \cite{2019AJ....158..166B,refId0}. Similar complications have been noted in other methods like the BCES-bisector \citep{2010arXiv1008.4686H}, thus we employ only the results obtained through the BCES-orthogonal method, which has been chosen as a favoured method in other applications (see e.g. \cite{Richard-Laferriere:2020mfo}).

Finally, the GRBs distance moduli, from the definition $\mu_{\mathrm{GRB}}=5 \log(d_L^{\mathrm{cal}}/\mathrm{Mpc}) +25 $, can be calculated by using all the fitting parameters obtained from the Amati relation for the sample of 74 GRBs with $0.117 \leq z \leq 5.283$. The variance of $\mu$ is computed by using error propagation method and is given by 
\begin{align}
\begin{split}
    \sigma_{\mu_{\mathrm{GRB}}}^2=&\left(\frac{\partial \mu_{\mathrm{GRB}}}{\partial A}\right)^2\sigma_{A}^2 + \left(\frac{\partial \mu_{\mathrm{GRB}}}{\partial B}\right)^2\sigma_{B}^2 +\\
    & 2\left(\frac{\partial \mu_{\mathrm{GRB}}}{\partial A}\right)\left(\frac{\partial \mu_{\mathrm{GRB}}}{\partial B}\right)\sigma_{AB} + \left(\frac{\partial \mu_{\mathrm{GRB}}}{\partial E_{\rm p}}\right)^2\sigma_{E_{\rm p}}^2\\
    & +\left(\frac{\partial \mu_{\mathrm{GRB}}}{\partial S_{\mathrm{bolo}}}\right)^2\sigma_{S_{\mathrm{bolo}}}^2,
\end{split}
\end{align}
where the covariances for $E_{\rm p}$ and $S_{\mathrm{bolo}}$ are absent since they are not correlated.

\begin{figure}
\begin{center}
\includegraphics[width=3.6in]{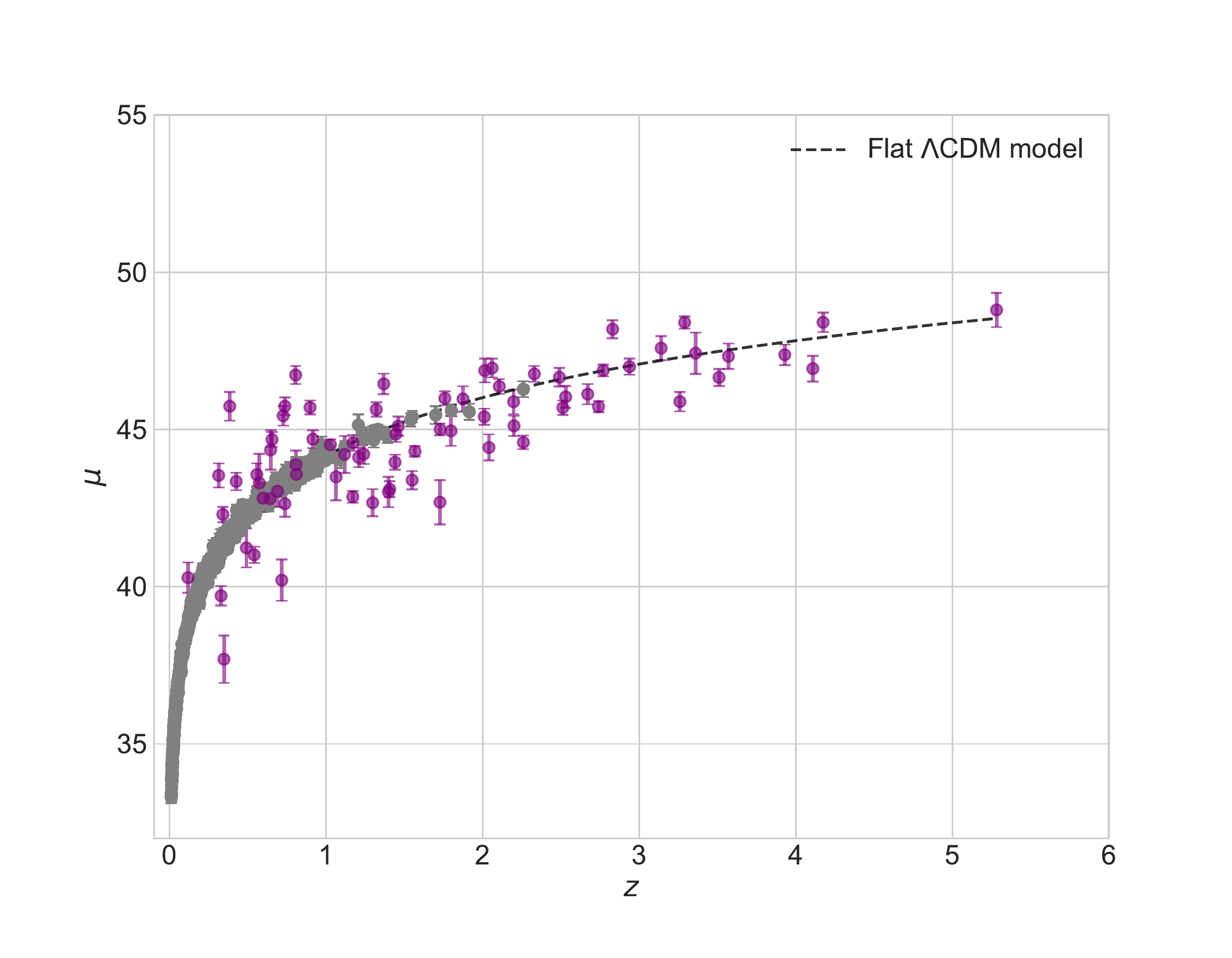}
\end{center}
\caption{Distance moduli  $\mu_{\mathrm{GRB}}$ for our 74 GRB  sample together with the SNe Ia distance moduli compared to the flat $\Lambda$CDM model with $\Omega_m=0.315$ and $h=0.674$.}
\label{fig:mu_vs_z}
\end{figure}

The distance moduli of the 74 GRBs, $\mu_{\mathrm{GRB}}$, and their 1$\sigma$ uncertainties calibrated through the Amati relation are listen in Table \ref{table:data_grbs}. The corresponding distribution of $\mu_{\mathrm{GRB}}$ versus $z$ is shown in Fig. \ref{fig:mu_vs_z} together the most recent compilation of SNe Ia, the Pantheon dataset \citep{Scolnic:2017caz}. 

\begin{table}
\scriptsize
\centering
\begin{tabular}{llll}
\hline
\hline
\addlinespace[0.1cm]
Name& $z_{\mathrm{GRB}}$ & $\mu_{\mathrm{GRB}}$ & $\sigma_{\mu_{\rm GRB}}$       \\
 \addlinespace[0.1cm]
\hline
\hline
\addlinespace[0.2cm]

 \text{GRB180728728} & 0.117 & 40.2876 & 0.485068 \\
 \text{GRB150727793} & 0.313 & 43.5373 & 0.38344 \\
 \text{GRB171010792} & 0.3285 & 39.7107 & 0.309248 \\
 \text{GRB130427324} & 0.34 & 42.2947 & 0.245225 \\
 \text{GRB130925173} & 0.347 & 37.6941 & 0.751191 \\
 \text{GRB140606133} & 0.384 & 45.7371 & 0.455169 \\
 \text{GRB190114873} & 0.425 & 43.3469 & 0.27912 \\
 \text{GRB091127976} & 0.4903 & 41.2349 & 0.620159 \\
 \text{GRB090618353} & 0.54 & 41.0103 & 0.261624 \\
 \text{GRB170607971} & 0.557 & 43.5656 & 0.358325 \\
 \text{GRB141004973} & 0.573 & 43.2978 & 0.923718 \\
 \text{GRB130215063} & 0.597 & 42.8165 & 0.442175 \\
 \text{GRB131231198} & 0.642 & 42.7956 & 0.213511 \\
 \text{GRB161129300} & 0.645 & 44.3482 & 0.630753 \\
 \text{GRB180720598} & 0.654 & 44.6778 & 0.239499 \\
 \text{GRB080916406} & 0.689 & 43.0348 & 0.497411 \\
 \text{GRB111228657} & 0.7163 & 40.2065 & 0.658414 \\
 \text{GRB140512814} & 0.725 & 45.4356 & 0.310337 \\
 \text{GRB160804065} & 0.736 & 42.6363 & 0.417647 \\
 \text{GRB090328401} & 0.736 & 45.7377 & 0.283133 \\
 \text{GRB100816026} & 0.8049 & 46.7315 & 0.288428 \\
 \text{GRB150514774} & 0.807 & 43.877 & 0.454466 \\
 \text{GRB151027166} & 0.81 & 43.5722 & 0.297084 \\
 \text{GRB091003191} & 0.8969 & 45.7004 & 0.218752 \\
 \text{GRB141225959} & 0.915 & 44.6923 & 0.28191 \\
 \text{GRB140508128} & 1.027 & 44.5137 & 0.172955 \\
 \text{GRB091208410} & 1.0633 & 43.4901 & 0.741808 \\
 \text{GRB180620660} & 1.1175 & 44.2046 & 0.589936 \\
 \text{GRB160509374} & 1.17 & 42.8561 & 0.186522 \\
 \text{GRB190324947} & 1.1715 & 44.5738 & 0.250219 \\
 \text{GRB140213807} & 1.2076 & 44.109 & 0.314352 \\
 \text{GRB090926914} & 1.24 & 44.214 & 0.312897 \\
 \text{GRB130420313} & 1.297 & 42.6682 & 0.430143 \\
 \text{GRB140801792} & 1.32 & 45.6352 & 0.225935 \\
 \text{GRB100414097} & 1.368 & 46.448 & 0.326704 \\
 \text{GRB100615083} & 1.398 & 43.0109 & 0.481797 \\
 \text{GRB160625945} & 1.406 & 43.0984 & 0.251807 \\
 \text{GRB100814160} & 1.44 & 43.957 & 0.239648 \\
 \text{GRB180314030} & 1.445 & 44.8533 & 0.253852 \\
 \text{GRB110213220} & 1.46 & 45.1038 & 0.305649 \\
 \text{GRB161117066} & 1.549 & 43.387 & 0.29187 \\
 \text{GRB100728095} & 1.567 & 44.3048 & 0.165154 \\
 \text{GRB100906576} & 1.727 & 42.6846 & 0.705178 \\
 \text{GRB120119170} & 1.728 & 45.004 & 0.184931 \\
 \text{GRB150314205} & 1.758 & 45.9886 & 0.220605 \\
 \text{GRB120326056} & 1.798 & 44.9544 & 0.475731 \\
 \text{GRB131011741} & 1.874 & 45.9691 & 0.404528 \\
 \text{GRB170705115} & 2.01 & 45.4027 & 0.259356 \\
 \text{GRB161017745} & 2.0127 & 46.8726 & 0.379906 \\
 \text{GRB140620219} & 2.04 & 44.4257 & 0.413362 \\
 \text{GRB150403913} & 2.06 & 46.9528 & 0.298582 \\
 \text{GRB090926181} & 2.1062 & 46.3678 & 0.234238 \\
 \text{GRB120624933} & 2.1974 & 45.88 & 0.392779 \\
 \text{GRB121128212} & 2.2 & 45.114 & 0.324018 \\
 \text{GRB081221681} & 2.26 & 44.5865 & 0.216956 \\
 \text{GRB141028455} & 2.33 & 46.7664 & 0.251315 \\
 \text{GRB130518580} & 2.49 & 46.6589 & 0.295154 \\
 \text{GRB081121858} & 2.512 & 45.6925 & 0.231653 \\
 \text{GRB170214649} & 2.53 & 46.0353 & 0.344863 \\
 \text{GRB120811649} & 2.671 & 46.1205 & 0.319856 \\
 \text{GRB140206304} & 2.74 & 45.7267 & 0.177794 \\
 \text{GRB081222204} & 2.77 & 46.8751 & 0.185688 \\
 \text{GRB110731465} & 2.83 & 48.1874 & 0.288535 \\
 \text{GRB181020792} & 2.938 & 46.9983 & 0.256644 \\
 \text{GRB140703026} & 3.14 & 47.5843 & 0.385486 \\
 \text{GRB140423356} & 3.26 & 45.8838 & 0.309091 \\
 \text{GRB140808038} & 3.29 & 48.4009 & 0.196729 \\
 \text{GRB110818860} & 3.36 & 47.4237 & 0.659063 \\
 \text{GRB170405777} & 3.51 & 46.6516 & 0.27214 \\
 \text{GRB090323002} & 3.57 & 47.3263 & 0.402843 \\
 \text{GRB120909070} & 3.93 & 47.3737 & 0.328148 \\
 \text{GRB090516353} & 4.109 & 46.9323 & 0.407857 \\
 \text{GRB120712571} & 4.1745 & 48.4067 & 0.310428 \\
 \text{GRB140304557} & 5.283 & 48.7986 & 0.542238 \\

\addlinespace[0.1cm]
 \hline
\end{tabular}
\caption{Distance moduli of 74 GRBs calibrated through the Amati relation.}
\label{table:data_grbs}
\end{table}

Before we proceed, we stress that, although  there is a debate as to whether the Amati relation is an intrinsic effect, or the result of detection biases, or even a combination of these two \citep{Butler2007,Li2007mnras,Ghirlanda2008mnras,Butler2009,Butler2010,Collazzi2012,Kocevski2012,Heussaff:2013sva,Petrosian2015}. Other works have claimed that the instrumental selection biases, even if they may affect the sample, cannot be responsible for the existence of the spectral-energy correlations \cite{Ghirlanda2008mnras,Nava2012MNRAS}. Additionally, the Amati relation could arguably change with redshift, but previous analyses have shown the opposite based on current data  (e.g., \cite{Demianski:2016zxi}). More thorough discussions supporting the reliability of Amati relation can be found in \cite{2017A&A...598A.112D,Dainotti:2017fhk,Dirirsa:2019fcs} 

\section{Dark Energy models} 
\label{models}

The dark energy dominance induces a period of accelerated expansion, well described by an effective equation of state $\omega_{\rm DE}=p_{\rm DE}/\rho_{\rm DE}$ (the label ${\rm DE}$ stands for dark energy). In the homogeneous and isotropic (FLRW) description of the Universe, a negative pressure (specifically $\omega_{\rm DE} < -1/3$) is required in order to induce the accelerated expansion. Furthermore, the parameter $\omega_{\rm DE}$ determines the gravitational properties of dark energy as well as its own evolution, derived from the energy-momentum conservation. The Friedmann equation (field equation) assuming a flat cosmology, pressureless matter, and negligible radiation is,
\begin{equation}
    \frac{H^2(z)}{H^2_0}= \Omega_m (1+z)^3 + \Omega_{\rm DE} \exp{\left(3 \int \frac{dz'}{1+z'}[1+\omega_{\rm DE}(z')]\right)},
    \label{DE_models}
\end{equation}

\noindent where the density fraction parameters are defined as $\Omega_m\equiv \rho_m(t_0)/\rho_c^0$ and $\Omega_{\rm DE}\equiv \rho_{\rm DE}(t_0)/\rho_c^0$ with critical density $\rho_c^0\equiv 3H_0^2/(8 \pi G)$. It is important to recall that in this work $H_0$ is not a parameter constrained by the Bayesian analysis. This  is because once we fit the Bezier curve, the value for the Hubble constant is set through the $\beta_0$ parameter (see Eq. (\ref{betas:2cases})).

From Eq. (\ref{DE_models}), we recover the expansion history for all the dark energy models here studied: 
\begin{enumerate}
\item $\Lambda$CDM model
    
In this case Eq. \eqref{DE_models} reads,
    \begin{equation}
    H^2(z)= H^2_0\left[\Omega_m (1+z)^3 + \Omega_{\rm DE}\right],
    \label{LCDM_model}
\end{equation}
where $\Omega_{\Lambda}\equiv \Omega_{\rm DE}$ is the density fraction due to a cosmological constant with equation of state $\omega_{\Lambda}=-1$. Requiring the consistency of Eq. \eqref{LCDM_model} at $z=0$, that is, $H(z=0)=H_0$, yields $\Omega_m+\Omega_{\Lambda}=1$. Thus, there is only one free parameter.

\item $\omega$CDM model
    
For the case in which $\omega_{\rm DE}$ is a constant such that $\omega_{\rm DE} \equiv \omega_0 \ne -1$, one gets
    \begin{equation}
    H^2(z)= H^2_0\left[\Omega_m (1+z)^3 + \Omega_{\rm DE}(1+z)^{3(1+\omega_{\rm DE})}\right],
    \label{wcdm_model}
\end{equation}
where $\Omega_{\rm DE}$ is the density fraction due to the dark energy fluid. Since flatness imposes the constraint $\Omega_m+\Omega_{\rm DE}=1$,  then the free parameters of this model are $\Omega_m$ and $\omega_0$.

\item CPL model
    
In the well-known CPL model, the equation of state is $\omega_{\rm DE} = \omega_0 + \omega_a \frac{z}{1+z}$, where $\omega_0$ and $\omega_a$ are both constants. The Hubble parameter equation, from Eq. \eqref{DE_models}, is 
    \begin{equation}
    H^2(z)= H^2_0\left[\Omega_m (1+z)^3 + \Omega_{\rm DE}(1+z)^{3(1+\omega_0+\omega_a)} \exp{\left( -\frac{3\omega_a z}{1+z}\right)}\right],
    \label{cpl_model}
\end{equation}
where $\Omega_{\rm DE}$ is the density fraction due to the dark energy fluid and again $\Omega_m+\Omega_{\rm DE}=1$ is imposed. Consequently, the free parameters of the model are $\Omega_m$, $\omega_0$ and $\omega_a$.

\end{enumerate}

Our task is to constrain the free parameters of each model by performing a Bayesian parameter estimation on the a set of up-to-date observations, including Type Ia Supernovae, BAO data and CMB data in the condensed form of shift parameters, complementing the GRB distance moduli reported here.

\section{Parameter estimation of Dark Energy Models} 
\label{methodology}

We have used the public Boltzmann code \textsc{class} \citep{Lesgourgues:2011re} to run the background evolution for all the dark energy models studied here. Then we use the cosmological parameter estimator \textsc{monte python} \citep{Audren:2012wb}, which is linked to \textsc{class} and adopts the Markov Chain Monte Carlo (MCMC) method to constrain the parameters of each DE model by fitting the cosmological data. The code employs the Metropolis-Hastings algorithm \citep{Metropolis:1953am,Hastings:1970aa} for sampling, and computes the Bayesian parameter inference of the posteriors with the convergence test given by the Gelman-Rubin criterion $R$ \citep{gelman1992}, where the condition $R-1<10^{-3}$ is imposed to end all of our chains. 

In addition to the calibrated samples of Gamma Ray Bursts listed above, the suite of datasets considered for our analysis includes those related to the expansion history of the universe (described by Eq.~\eqref{DE_models}), i.e., the ones describing the distance-redshift relations as detailed below.

\subsection{Observational data} 
\label{data}
\subsubsection{Type Ia Supernovae (SNe Ia)}

One of the latest SNe Ia data compilation is the Pantheon sample \citep{Scolnic:2017caz} which consists of 1048 SNe with the redshift spanning $0.01<z < 2.3$. This sample is a combination of 365 spectroscopically confirmed SNe Ia discovered by the Pan-STARRS1(PS1) Medium Deep Survey together with the subset of 279 PS1 SNe Ia ($0.03 < z < 0.68$) with distance estimates from SDSS, SNLS and several low-$z$ and Hubble Space Telescope samples (see Table 4 of \cite{Scolnic:2017caz}). In order to 
{include this data, we define} 
\begin{equation}
\chi^2_{SN}= \mathrm{\Delta} \mathrm{\mu^{SN}} \cdot \mathrm{C_{SN}^{-1}} \cdot \mathrm{\Delta} \mathrm{\mu^{SN}},
\label{Eq:chiSNIa}
\end{equation}
where $\mathrm{C_{SN}}$ is the full systematic covariance matrix and $\mathrm{\Delta} \mathrm{\mu^{SN}}=\mathrm{\mu_{theo}}
-\mathrm{\mu_{obs}}$ is the vector of the differences between the observed and theoretical value of the distance modulus, $\mathrm{\mu}$, with the absolute magnitude taken as nuisance parameter.

\subsubsection{Baryon Acoustic Oscillations (BAO)}

\begin{table}
\centering
\begin{tabular}{lcc}
\hline
\hline
\addlinespace[0.1cm]

Survey& $z_{\mathrm{BAO}}$ & Measurement       \\
 \addlinespace[0.1cm]
\hline
\hline
\addlinespace[0.2cm]
6DF & $0.106$ & $r_{\mathrm{s}}/D_V$    \\
\addlinespace[0.1cm]
SDSS DR7 MGS &$0.15$ & $D_V/r_{\mathrm{s}}$   \\
\addlinespace[0.1cm]
SDSS DR12 galaxies &$0.38$, $0.51$, $0.61$  & $ D_{A}/r_{\mathrm{s}}$, $ D_H/r_{\mathrm{s}}$   \\
\addlinespace[0.1cm]
 \hline
\end{tabular}
\caption{BAO measurements from various surveys, \protect\cite{2011MNRAS.416.3017B,Ross:2014qpa,Alam:2016hwk}, adopted in this work. }
\label{table:data_bao}
\end{table}

We used the low redshift galaxy BAO data listed in Table \ref{table:data_bao}. The data provide measurements of three types of ratios of comoving distance: the angular scale of the BAO ($D_{A}(z)/r_\mathrm{s}$), the redshift-space BAO scale ($D_H(z)/r_\mathrm{s}$), with $D_H(z)=c/H(z)$~ \citep{Alam:2016hwk}, and the spherically-averaged BAO scale ($D_V(z)/r_\mathrm{s}$) \citep{2011MNRAS.416.3017B,Ross:2014qpa}. Here $r_\mathrm{s}$ is the comoving sound horizon at the end of the baryon drag epoch, given by 
\begin{equation}
    r_\mathrm{s}=\int_{z_d}^\infty \frac{c_s(z)}{H(z)}dz,
\end{equation}
where $c_s$ denotes the sound speed in the primordial photon-baryon plasma given by  $c_s=3^{-1/2}c [1+\frac{3}{4}\rho_b(z)/\rho_{\gamma}(z)]^{-1/2}$. $D_A(z)$ is the comoving angular diameter distance 
\begin{equation}
    D_A(z)=c\int_{0}^z \frac{dz'}{H(z')},
\end{equation}
and $D_V(z)$ is the spherically averaged combination of transverse and radial BAO modes, 
\begin{equation}
    D_V(z) = \left[zD_H(z)D_A^2(z) \right]^{1/3}.
\end{equation}

Thus, the corresponding $\chi_{BAO}^2$ for BAO data is given by
\begin{equation}
\chi^2_{BAO}= \mathrm{\Delta} \mathbfcal{F}^{BAO} \cdot \mathbf{C}_{BAO}^{-1} \cdot \mathrm{\Delta} \mathbfcal{F}^{BAO},
\label{Eq:chiBAO}
\end{equation}
where $\mathrm{\Delta} \mathbfcal{F}^{BAO}=\mathcal{F}_{theo}-\mathcal{F}_{obs}$ is the difference between the observed and theoretical value of the observable quantity for BAO, which may vary from one survey to another, and $\mathbf{C}^{-1}_{BAO}$ is the respective inverse covariance matrix.

\subsubsection{Cosmic Microwave Background (CMB)}

Instead of the full data of the CMB anisotropies, we use CMB data in the condensed form of shift parameters reported by \cite{Chen:2018dbv}, derived from the last release of the Planck results \citep{Aghanim:2018eyx}. Evidently, the analysis proceeds much faster in this way than by performing an analysis involving the full CMB likelihood. 

As argued by several authors, the shift parameters $(R,l_A,\Omega_bh^2,n_s)$ provide an efficient summary of CMB data as far as DE constraints are concerned \citep{Kosowsky:2002zt,Wang:2007mza,Mukherjee:2008kd,Ade:2015rim}. This set of parameters can be used to study models with either non-zero curvature or a smooth DE component, as is our case, but not with modifications of gravity \citep{Mukherjee:2008kd,Ade:2015rim}.

The first two quantities in the vector $(R,l_A,\Omega_bh^2,n_s)$ are defined as 
\begin{equation}
R \equiv \sqrt{\Omega_m H_0^2} \frac{r(z_{*})}{c},
\end{equation}
\begin{equation}
l_A\equiv \pi \frac{r(z_{*})}{r_s(z_{*})},
\end{equation}
where $r(z)$ is the comoving distance, here evaluated at photon-decoupling epoch $z_{*}$. 
The corresponding $\chi^2$ for the CMB is thus
\begin{equation}
\chi^2_{CMB}= \mathrm{\Delta} \mathbfcal{F}^{CMB} \cdot \mathbf{C}_{CMB}^{-1} \cdot \mathrm{\Delta} \mathbfcal{F}^{CMB}    ,
\label{Eq:chiCMB}
\end{equation}
where $\mathcal{F}^{CMB}=(R,l_A,\Omega_bh^2,n_s)$ is the vector of the shift parameters and $\mathbf{C}^{-1}_{CMB}$ is the respective inverse covariance matrix. The mean values for these shift parameters as well as their standard deviations and normalized covariance matrix are taken from Table 1 of \cite{Chen:2018dbv}.

\subsubsection{Gamma-Ray Bursts (GRBs)}

We test our method by analysing two samples. The first one consists on 193 GRBs calibrated in \cite{Amati:2018tso} which cover the redshift range $0.03351 \leq z \leq 8.1$. The second sample is our set of 74 Fermi-GRBs, a selection described in detail above, with a redshift range $0.117 \leq z \leq 5.283$.

The $\chi^2$ function for the GRBs data is defined analogous to Eq. \eqref{Eq:chiSNIa} of SNe Ia data, that is, 
\begin{equation}
\chi^2_{GRB}= \mathrm{\Delta} \mathrm{\mu}^{GRB} \cdot \mathrm{C^{-1}}_{GRB} \cdot \mathrm{\Delta} \mathrm{\mu}^{GRB},
\label{Eq:CSNIa}
\end{equation}
where $\mathrm{C}_{GRB}$ is a diagonal matrix containing $\sigma^2_{\mu}$ and $\mathrm{\Delta} \mathrm{\mu}^{GRB}=\mathrm{\mu_{theo}}
-\mathrm{\mu_{estimated}}$ is the vector of the differences between the theoretical and \textit{estimated} value of the distance moduli for the GRBs.

\section{RESULTS} \label{Results}

We have obtained the constraints for the $\Lambda$CDM, $\omega$CDM and CPL models from the latest observational data of SNe Ia, BAO, CMB distance priors inferred from the final Planck 2018 data, and including either the 193 GRBs calibrated by \cite{Amati:2018tso} labeled as sample (1) or the 74 GRBs calibrated in this work labeled as sample (2-nosys) or sample (2-sys). For comparison, SNe Ia + BAO + CMB without GRBs have been also analysed in order to highlight the contribution of GRBs to the joint cosmological constraints. 

We used flat priors for the free parameters of the models analyzed with ranges set as follow: $\Omega_{m} \in (0.1, 0.5)$ (all models). $\omega_{0} $ in the $\mathrm{\omega CDM}$ model with no priors. Finally, the CPL parameters are allowed to take values in the ranges $\omega_{0} \in (-2, 0),~ \omega_{a} \in (-5, 5) $.  
Note that if we wanted to constraint the value for the Hubble constant $H_0$ from all cosmological data employed in this work, it would be necessary to leave $\beta_0$ as a free parameter before and after the Bezier function fitting.

\setlength\dashlinedash{0.2pt}
\setlength\dashlinegap{1.5pt}
\setlength\arrayrulewidth{0.3pt}

\begin{table*}
\centering
\begin{tabular}{lcrcrcrcr}
\hline
\hline
\addlinespace[0.1cm]
\multirow{2}[3]{*}{\textbf{ }} & \multicolumn{2}{c}{\small{\textbf{SNIa+BAO+CMB}}} & \multicolumn{2}{c}{\small{\textbf{SNIa+BAO+CMB + GRBs(1)}}} & 
\multicolumn{2}{c}{\small{\textbf{ALL + GRBs(2-no sys)}}} & 
\multicolumn{2}{c}{\small{\textbf{ALL + GRBs(2-sys)}}} \\
\cmidrule(lr){2-3} \cmidrule(lr){4-5} \cmidrule(lr){6-7} \cmidrule(lr){8-9} 
 &  best-fit & mean$\pm\sigma$ &  best-fit & mean$\pm\sigma$ & best-fit & mean$\pm\sigma$ & best-fit & mean$\pm\sigma$  \\
 \addlinespace[0.1cm]
\hline
\hline
\addlinespace[0.2cm]
$\Omega_{\mathrm{m}}$ &$0.3179$ & $0.3179_{-0.00058}^{+0.00057}$ & $0.3180$ & $0.3180_{-0.00057}^{+0.00057}$ & $0.3180$ & $0.3180_{-0.00057}^{+0.00057}$ &$0.3180$ & $0.3180_{-0.00057}^{+0.00057}$ \\
\addlinespace[0.2cm]
 \hline
\end{tabular}
\caption{Constraints at 68$\%$ C.L. on the cosmological parameter $\Omega_\mathrm{m}$ in case of the $\Lambda$CDM model using different combinations of datasets. 1) The joint analysis of the SNe Ia, BAO and CMB distance priors, 2) SNe Ia + BAO + CMB + the sample of 193 GRBs calibrated by~\protect\cite{Amati:2018tso}, labeled as GRBs(1). For 3) and 4)  SNe Ia + BAO + CMB is condensed as ALL and the sample of 74 GRBs from Fermi-GBM catalog calibrated in this work, labeled as GRBs(2-no sys) and GRBs(2-sys) without and with extra bias in the CC Hubble data, respectively, is added.}
\label{table:lcdm}
\end{table*}

In Table \ref{table:lcdm}, the best-fit values for $\Omega_m$ with 1-$\sigma$ uncertainties are shown. In spite of the relatively large dispersion observed in our calibrated GRBs sample, the results obtained by using either GRBs(2-nosys) or GRBs(2-sys) plus SNIa+BAO+CMB are in good agreement with the results we obtained by using the calibrated sample by \cite{Amati:2018tso}, namely the results obtained by using SNIa+BAO+CMB+GRBs(1), and these are consistent with the Planck 2018 results  \citep{Aghanim:2018eyx} at 1$\sigma$ confidence level\footnote{Note that our results, either including GRBs(1), GRBs(2-nosys) or GRBs(2-sys), are  consistent with the ones reported by \cite{Amati:2018tso} only at 2$\sigma$ on  $\Omega_m$ for this model.  This in part may be attributed to the suite of data used there, which included JLA SNe Ia \citep{Betoule:2014frx} and no CMB or BAO data.}.

\begin{figure}
\begin{center}
\includegraphics[width=3.3in]{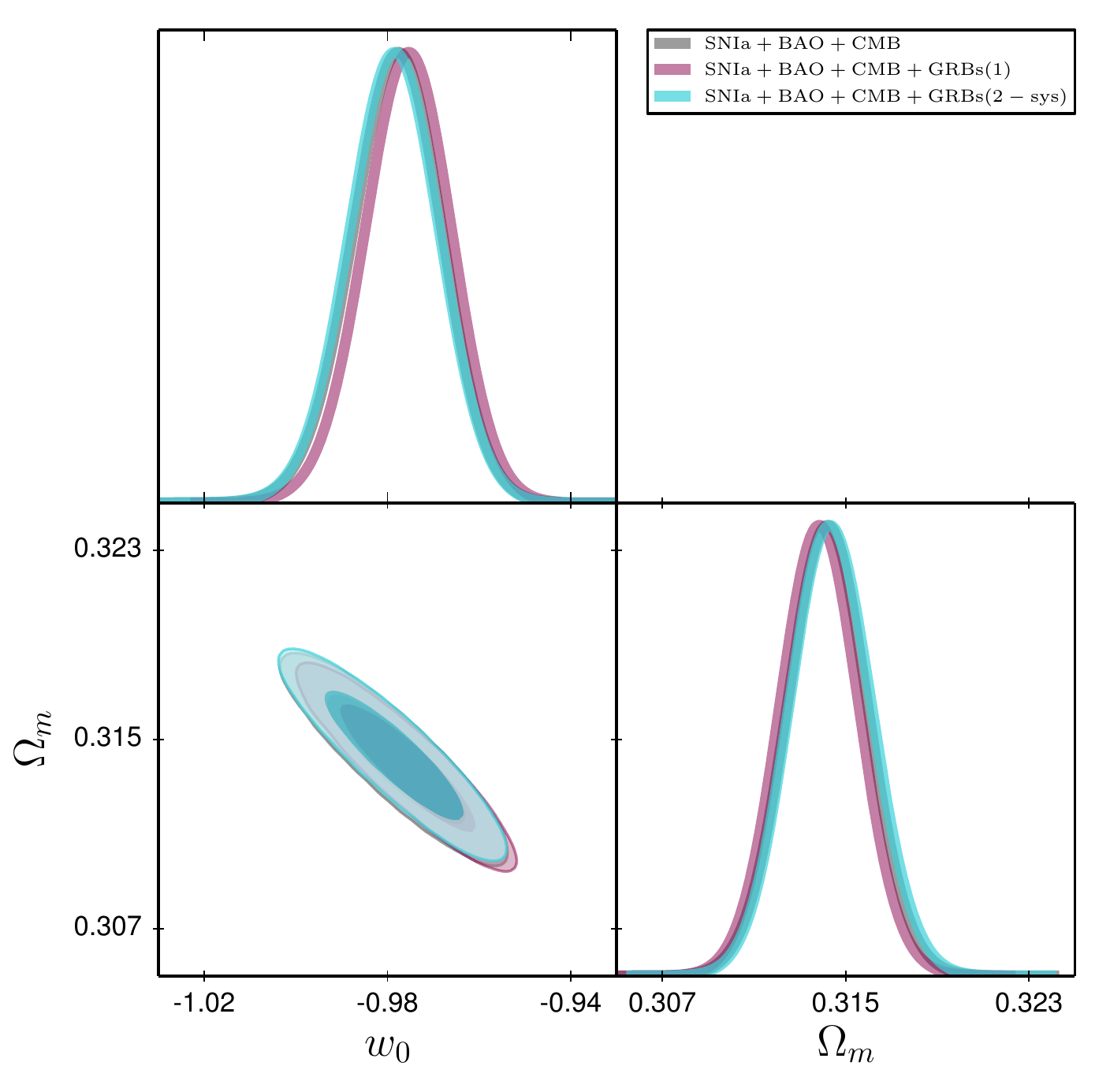}
\end{center}
\caption{Constraints at the 68$\%$ and 95$\%$ C.L. on the ($\omega_0$, $\Omega_{\mathrm{m}}$) plane from the combinations of datasets 1) The joint analysis of the SNe Ia, BAO and CMB distance priors, 2) SNe Ia + BAO + CMB + the sample of 193 GRBs calibrated by~\protect\cite{Amati:2018tso}, labeled as GRBs(1) and 3)  SNe Ia + BAO + CMB + the sample of 74 GRBs from Fermi-GBM catalog presented in this work taking into account extra bias due to the systematic uncertainties in the CC Hubble dataset, labeled as GRBs(2-sys). } 
\label{fig:wcdm}
\end{figure}

\begin{table*}
\centering
\begin{tabular}{lcrcrcrcrcr}
\hline
\hline
\addlinespace[0.1cm]
\multirow{2}[3]{*}{\textbf{ }} & \multicolumn{2}{c}{\small{\textbf{SNe Ia + BAO + CMB}}} & \multicolumn{2}{c}{\small{\textbf{SNe Ia + BAO + CMB + GRBs(1)}}} & 
\multicolumn{2}{c}{\small{\textbf{ALL + GRBs (2- no sys)}}} & \multicolumn{2}{c}{\small{\textbf{ALL + GRBs (2- sys)}}}\\
\cmidrule(lr){2-3} \cmidrule(lr){4-5} \cmidrule(lr){6-7} \cmidrule(lr){8-9} 
 &  best-fit & mean$\pm\sigma$ &  best-fit & mean$\pm\sigma$ & best-fit & mean$\pm\sigma$ & best-fit & mean$\pm\sigma$ \\
 \addlinespace[0.1cm]
\hline
\hline
\addlinespace[0.2cm]
$\omega_{0}$ & $-0.9776$ & $-0.9782_{-0.0096}^{+0.01}$ &
$-0.9755$ & $-0.9757_{-0.0096}^{+0.0099}$ & 
$-0.9783$ & $-0.9785_{-0.0097}^{+0.01}$ &
$-0.9779$ & $-0.9786_{-0.0098}^{+0.01}$    \\
\addlinespace[0.2cm]
$\Omega_{\mathrm{m}}$ &$0.3141$ & $0.3142_{-0.0018}^{+0.0018}$&
$0.3138$ & $0.3138_{-0.0018}^{+0.0018}$ & 
$0.3143$ & $0.3143_{-0.0018}^{+0.0018}$ &
$0.3142$ & $0.3144_{-0.0018}^{+0.0018}$  \\
\addlinespace[0.2cm]
 \hline
\end{tabular}
\caption{Constraints at 68$\%$ C.L. on the cosmological parameters of the $\omega$CDM model using the combinations of datasets mentioned in Table \ref{table:lcdm}. }
\label{table:wcdm}
\end{table*}

The results of a similar analysis for the two parameters of the $\omega$CDM model are displayed in Figure \ref{fig:wcdm}. We show only the results from the calibrated sample of GRBs that comes from the inclusion of extra bias in the CC Hubble dataset. In this case our calibrated sample yields a value for the $\omega_0$ parameter consistent with the ones obtained from the other datasets employed here.

The corresponding best-fits of the analysis are listed in Table \ref{table:wcdm}. It is important to note that the parameter estimation of \cite{Amati:2018tso} for this model yields best fit values for $\omega_0$  that fall far from our fit of their same dataset and show a much larger uncertainty than ours.

This advocates for our method as an improvement over previous analyses of GRB data.
\begin{figure}
\begin{center}
\includegraphics[width=3.3in]{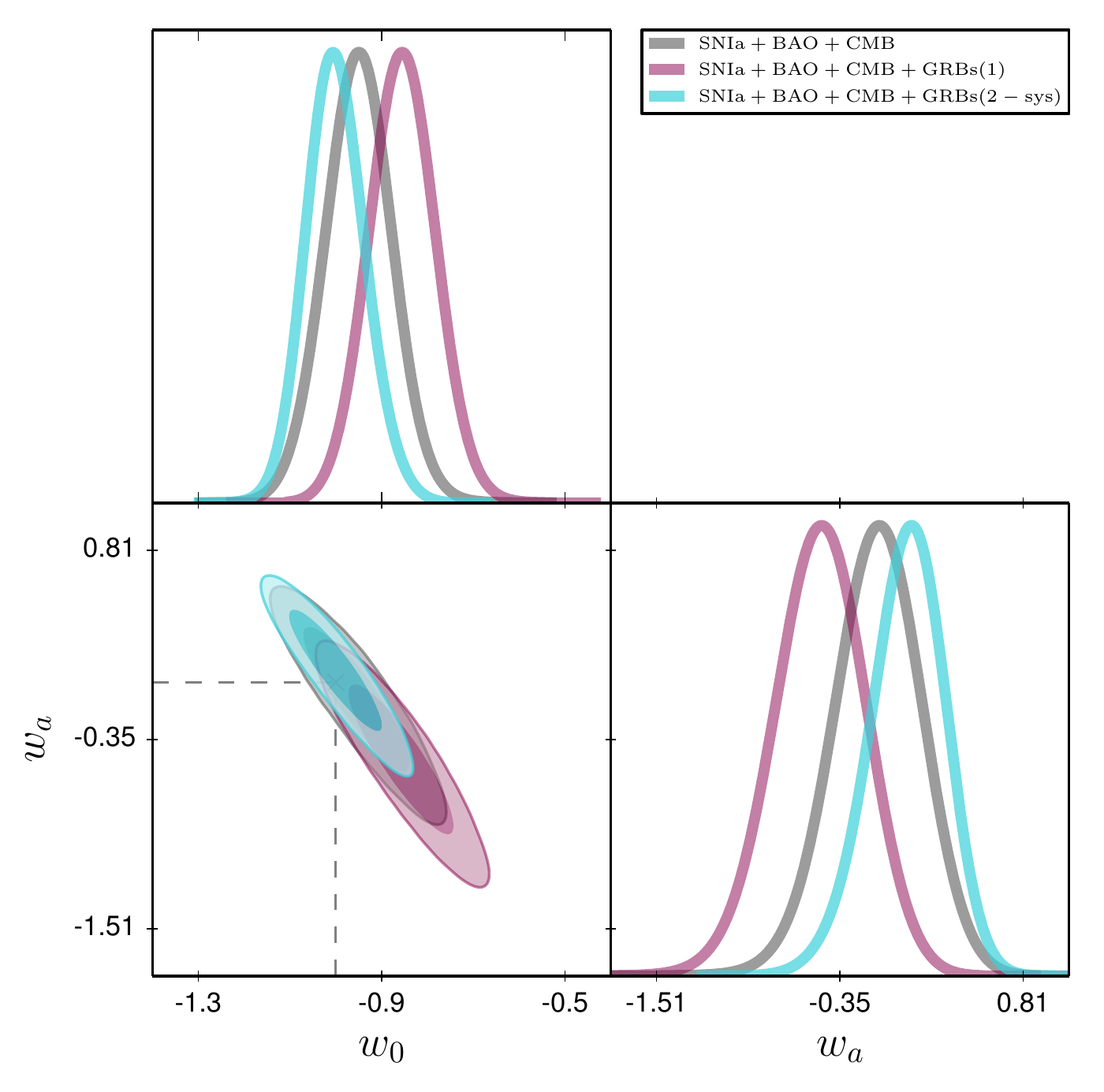}
\end{center}
\caption{Constraints at the 68$\%$ and 95$\%$ C.L. on the ($\omega_0$, $\omega_{a}$)  plane from the combinations of datasets mentioned in Fig. \ref{fig:wcdm}. }
\label{fig:cpl}
\end{figure}

\begin{table*}
\centering
\begin{tabular}{lcrcrcrcr}
\hline
\hline
\addlinespace[0.1cm]
\multirow{2}[3]{*}{\textbf{ }} & \multicolumn{2}{c}{\small{\textbf{SNe Ia + BAO + CMB}}} & \multicolumn{2}{c}{\small{\textbf{SNe Ia + BAO + CMB + GRBs(1)}}} & 
\multicolumn{2}{c}{\small{\textbf{ALL + GRBs(2-no sys)}}} & 
\multicolumn{2}{c}{\small{\textbf{ALL + GRBs(2- sys)}}} \\
\cmidrule(lr){2-3} \cmidrule(lr){4-5} \cmidrule(lr){6-7} \cmidrule(lr){8-9}
 &  best-fit & mean$\pm\sigma$ &  best-fit & mean$\pm\sigma$ & best-fit & mean$\pm\sigma$ & best-fit & mean$\pm\sigma$  \\
 \addlinespace[0.1cm]
\hline
\hline
\addlinespace[0.2cm]
$\omega_{0}$ & $-0.9631$ & $-0.9515_{-0.078}^{+0.074}$  &
$-0.8628$ & $-0.8548_{-0.075}^{+0.073}$   &
$-1.002$ & $-0.9933_{-0.069}^{+0.066}$ &
$-1.009$ & $-0.9982_{-0.068}^{+0.064}$   \\
\addlinespace[0.2cm]
$\omega_{a}$ &$-0.05507$ & $-0.1079_{-0.27}^{+0.3}$ & 
$-0.4361$ & $-0.4782_{-0.28}^{+0.31}$  &
$0.09276$ & $0.05013_{-0.23}^{+0.27}$ &
$0.1135$ & $0.06781_{-0.22}^{+0.26}$  \\
\addlinespace[0.2cm] 
$\Omega_{\mathrm{m}}$ &$0.3142$ & $0.3144_{-0.0019}^{+0.0018}$  &
$0.3144$ & $0.3147_{-0.0019}^{+0.0018}$  &
$0.3139$ & $0.3142_{-0.0019}^{+0.0018}$ &
$0.3141$ & $0.3143_{-0.0019}^{+0.0018}$   \\

\addlinespace[0.1cm] 
 \hline
\end{tabular}
\caption{Constraints at 68$\%$ C.L. on the cosmological parameters of the CPL model using the combinations of datasets mentioned in Table \ref{table:lcdm}.}
\label{table:cpl}
\end{table*}

 In Figure \ref{fig:cpl} we show 1$\sigma$ (dark colours) and 2$\sigma$ (light colours) error contours in the ($\omega_0$, $\omega_{a}$) parameter-space of the CPL model, resulting from the combination of samples indicated above. 
 The  datasets tested for this model show consistency in parameter values at $1\sigma$.
Furthermore, the parameter values fitting the new sample of GRBs favours the $\Lambda$CDM model more than the other two cases. This is clear from Table \ref{table:cpl}, where the $\Lambda$CDM values $(\omega_0=-1, \omega_a = 0)$ lie close to the central values of our new GRBs dataset.

We also note that the values of the parameter $\omega_0$ in the CPL model from the three datasets employed in this work, intersect the range of values resulting from the Planck 2018 data at 1$\sigma$ \citep{Aghanim:2018eyx},  in contrast with the recent results of \cite{Demianski:2019vzl} (where a sample of GRBs covering a redshift range of $0.033\leq z< 9$ is employed, together with direct $H(z)$ measurements \citep{Farooq:2013hq} and the past compilation of SNe Ia Union 2.1 \citep{Union21}). We emphasise that our adequate handling of errors, and a suitable choice of calibration method, yield a significantly tighter confidence region for the parameters of the models here presented, in comparison to the posteriors of previous works.

For a quick insight into model comparison, we compute the Bayesian information criterion (BIC) introduced by \citet{Schwarz:1978tpv}, defined as  
\begin{equation}
BIC = -2 \ln{\mathcal{L_{\mathrm{max}}}}+ k \ln{N},    
\end{equation}
where $N$ is the number of datapoints used in the fit and $k$ is the number of free parameters of the cosmological model. \footnote{It is worth mentioning that the BIC criterion, as well as the AIC (Akaike Information Criterion) \citep{akaike1974new}, the other common method used to perform model selection, are in fact approximate selection tools. 
A formal model comparison can be achieved through Bayesian Evidence, which is widely known as the most reliable statistical tool for model comparison. See \cite{Cedeno:2019cgr}, for a recent work in this direction, and more details of Bayesian Model Selection in \cite{Trotta:2008qt}.} 
In practice, one computes BIC for each of the candidate models and selects the model of reference to calculate the difference $\Delta \mathrm{BIC}= \Delta \chi^2 + \Delta k \ln N$. A difference in $\Delta \mathrm{BIC} $ of 2 is considerable positive evidence against the model with higher BIC, while a $\Delta \mathrm{BIC} $ of 6 is considered to be strong evidence \cite{Jeffreys1961}. For the model comparison we choose the $\Lambda$CDM model as the model of reference because it shows lowest BIC score. Thus, for the case in which our calibrated sample of GRBs is coming from CC Hubble data with extra systematics uncertainties, we obtained $\Delta \mathrm{BIC}_{\mathrm{\omega CDM}}= 2.25$ and $\Delta \mathrm{BIC}_{\mathrm{CPL}}= 9.08$ indicating positive evidence against $\omega$CDM model and strong evidence against the CPL model. The evidence against alternative models is slightly weaker when our GRBs sample is removed, in this case we found  $\Delta \mathrm{BIC}_{\mathrm{\omega CDM}}= 1.97$ and $\Delta \mathrm{BIC}_{\mathrm{CPL}}= 8.88$ showing a positive weak evidence against the $\omega$CDM but confirming the strong evidence against the CPL model. Thus, by employing the BIC selection criteria to compare the statistical performance of all the studied models with and without our calibrated sample of GRBs, we found that $\Lambda$CDM model is still preferred with respect to the $\omega$CDM and CPL models.

\begin{figure}
\begin{center}
\includegraphics[width=3.7in]{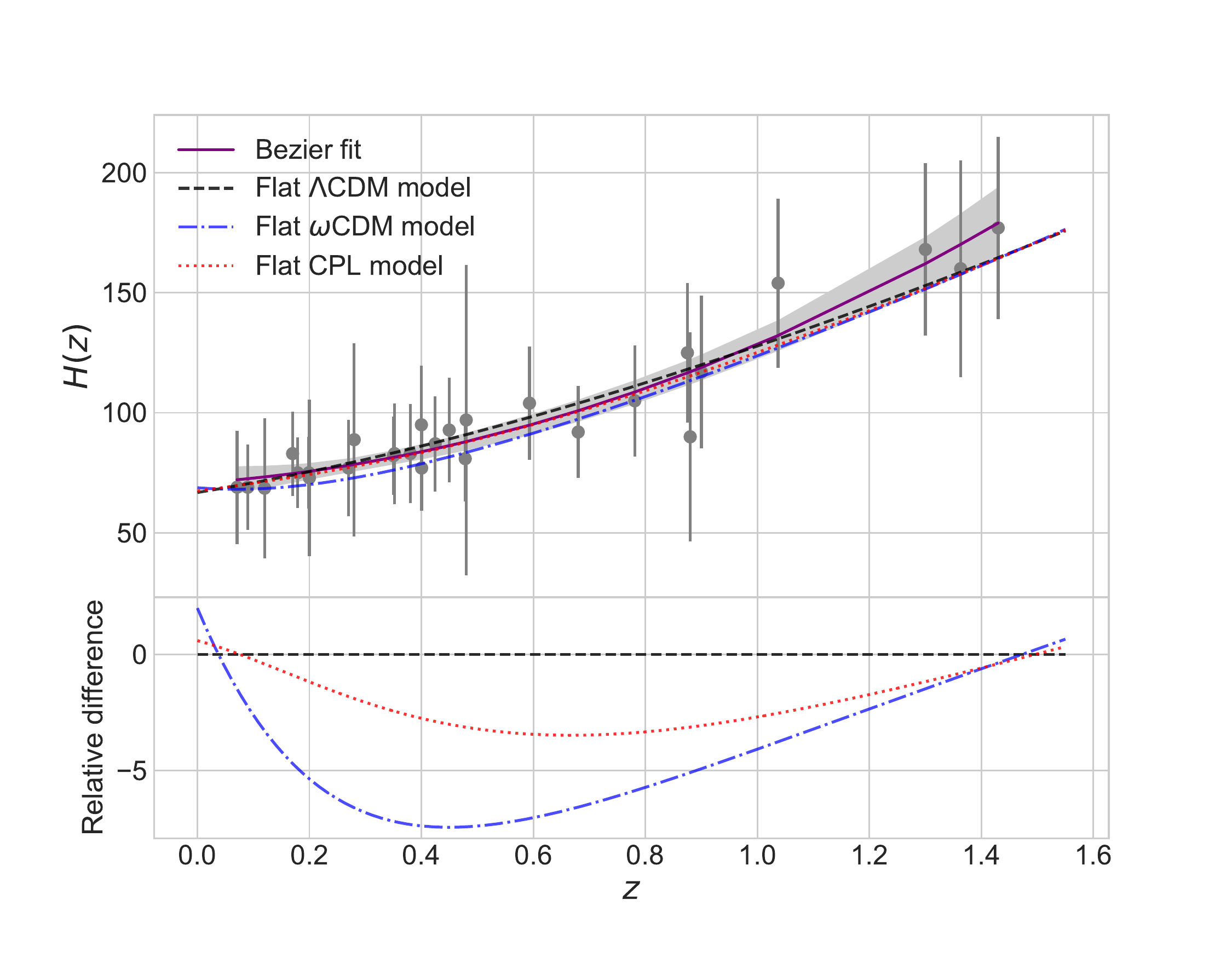}
\end{center}
\caption{In the upper panel, the 28 Hubble parameter datapoints with the fit to the B\'ezier curve and its respective 1$\sigma$ confidence region (shaded region including the maximum uncertainty at 1-$\sigma$ from the recent treatment of uncertainties from SPS modelling, stellar library, and stellar metallicity \citep{Moresco:2020fbm}.
Additionally, the fitted Hubble function for each of the cosmological models constrained in this paper are shown. In the lower panel, the relative differences of the models w.r.t. the $\Lambda$CDM one.}
\label{fig:BezierCurveModels}
\end{figure}


\section{Conclusions and Discussion}
\label{sec:conclusions}

Through the criteria described in Sec. \ref{GRBs} we have carefuly selected a sample of 74 GRBs as tracers of the luminosity distance. The Amati relation for these GRBs is calibrated in a model-independent way. Such calibration, described in Section \ref{Calibration}, relies on $H(z)$ data coming from the CC approach as our calibration source at low redshifts. 

We have computed and incorporated our GRB distance moduli to a suite of observations complemented by the latest CMB, BAO and SNe 1a data in order to fit parameters of Dark Energy and test the usefulness of our sample. We find consistency with previous works for $\Lambda$CDM and $\omega$CDM models at 1$\sigma$ in the posterior contours of the relevant parameters, with the bonus of a much tighter confidence region for the parameters. This is particularly evident for the CPL model, where the preferred values are close to those of $\Lambda$CDM, in contrast to recent results in the literature \citep{Demianski:2019vzl}

Since the parameter values of alternative models and the BIC criterion suggest preference of a $\Lambda$CDM cosmology, we argue that this is still the better model to describe all of the data.

There is room for improvement in the analysis here presented. Most prominently, the Bezier fit to the CC Hubble data is not reproduced by the $H(z)$ function of the cosmological models as shown in Figure~\ref{fig:BezierCurveModels}. This means that the cosmology dictated by the CC Hubble data is not reproduced in the posterior values of the tested models. As an alternative proposal to circumvent the circularity problem, one might find convergence by jointly calibrating the GRB Amati relation and the fitting to the cosmological model, including the CC Hubble data as one more set of observations \citep{Firmani:2006zg,Firmani:2005gs,Xu:2005uv,Schaefer:2006pa,Li:2007mt}. We shall perform these alternative tests and compare strategies to tackle the circularity problem in a follow-up study.

In the meantime, given the reduction of the confidence region in the parameter space of alternative models obtained from our GRB sample and method, we argue in favour of our analysis when considering GRBs and other luminosity distance probes.



\section*{Acknowledgements}
A.M. acknowledges support from postdoctoral grants from DGAPA-UNAM. The authors acknowledge sponsorship from CONACyT through grant CB-2016-282569. We also acknowledge the publicly available data from Fermi collaboration. The authors are grateful to the anonymous referee for his/her insightful comments and suggestions that helped improving the initial manuscript.


\section*{Data availability}

The data underlying this article are available in the article.




\bibliography{biblio}
\bibliographystyle{mnras}






\bsp	
\label{lastpage}
\end{document}